\documentclass{emulateapj}
\usepackage{epsf}
\usepackage{graphicx}
\usepackage{float}
\usepackage{afterpage}
\usepackage{relsize}
\usepackage{amsmath}
\usepackage{epsfig}
\usepackage{enumitem}
\usepackage{subfigure}

\newcommand{\ba}{\begin{eqnarray}}
\newcommand{\ea}{\end{eqnarray}}

\begin{document}

\title{Birth Locations of the Kepler Circumbinary Planets}

\author{Kedron Silsbee\altaffilmark{1} \& Roman R. Rafikov\altaffilmark{1} }
\altaffiltext{1}{Department of Astrophysical Sciences, 
Princeton University, Ivy Lane, Princeton, NJ 08540; 
ksilsbee@astro.princeton.edu}

\begin{abstract}
The {\it Kepler} mission has discovered about a dozen circumbinary planetary systems, all containing planets on $\sim$ 1 AU orbits. We place bounds on the locations in the circumbinary protoplanetary disk, where these planets could have formed through collisional agglomeration starting from small (km-sized or less) planetesimals.  We first present a model of secular planetesimal dynamics that accounts for the (1) perturbation due to the eccentric precessing binary, as well as the (2) gravity and (3) gas drag from a precessing eccentric disk. Their simultaneous action leads to rich dynamics, with (multiple) secular resonances emerging in the disk. We derive analytic results for size-dependent planetesimal eccentricity, and demonstrate the key role of the disk gravity for circumbinary dynamics. We then combine these results with a simple model for collisional outcomes and find that in systems like Kepler 16, planetesimal growth starting with 10-100 m planetesimals is possible outside a few AU. The exact location exterior to which this happens is sensitive to disk eccentricity, density and precession rate, as well as to the size of the first generation of planetesimals. Strong perturbations from the binary in the inner part of the disk, combined with a secular resonance at a few AU inhibit the growth of km-sized planetesimals within $2-4$ AU of the binary. In situ planetesimal growth in the Kepler circumbinary systems is possible only starting from large (few-km-sized) bodies in a low-mass disk with surface density $\lesssim 500$ g cm$^{-2}$ at 1 AU. 
\end{abstract}


%%%%%%%%%%%%%%%%%%%%%%%%%%%%%%%%%%%%%%%%%%%%%
%%%%%%%%%%%%%%%%%%%%%%%%%%%%%%%%%%%%%%%%%%%%%

\section{Introduction}
\label{sect:intro}

%%%%%%%%%%%%%%%%%%%%%%%%%%%%%%%%%%%%%%%%%%%%%

Recent discoveries of exoplanets in stellar binaries by radial velocity surveys and the {\it Kepler} mission stimulated significant interest in understanding the origin of such planetary systems. Planet-hosting stellar binaries come in two flavors, commonly denoted as S-type or P-type \citep{DVOR82}.  They correspond to systems where the planet orbits one star with the other as an external perturber (S-type), or where the planet is in orbit around both stars (P-type).  

In this paper, we focus on P-type or circumbinary systems.  {\it Kepler} has so far revealed to us eight such systems, all containing sub-Jovian planets in orbits around main-sequence binaries. These systems have a range of parameters, but generally the stars have masses $\sim M_\odot$, and are on moderately eccentric orbits with semi-major axes $a_b\sim 0.1-0.2$ AU. The planetary orbits have semi-major axes smaller than $1.1$ AU.  The parameters of known {\it Kepler} circumbinary planets are summarized in Table \ref{table:sysParams}.

There may also exist another population of circumbinary planets hinted at by transit timing of post-common envelope binaries.  The most plausible of these are two planets around the NN Serpentis binary system \citep{Marsh14}.  It has been suggested \citep{Volschow14} that such planets are not primordial and formed from matter ejected during common envelope evolution. There are also a few directly imaged long period planetary mass circumbinary companions at projected separations $\gtrsim 100$ AU from the host star, e.g. ROXs 42Bb \citep{Currie14}. We will not address the origin of such systems in this work. 

Planet formation in a circumbinary disk faces challenges due to vigorous planetesimal excitation driven by the non-Newtonian potential of the binary.  Large relative velocities between planetesimals are harmful to coagulation because collisions lead to planetesimal destruction rather than merging.  It is generally believed that the circumbinary planets could not have formed in situ via collisional agglomeration \citep{Meschiari2012, Paardekooper2012, Marzari13}, due to the high collision speeds of km sized planetesimals.  That said, \citet{Meschiari2014} proposes a model for in situ formation, in which  large planetesimals form at the pressure maximum near the inner edge of the disk where the surface density of solids is expected to be increased. \citet{Kenyon2015} suggest that circumbinary planetesimals may settle onto special class of orbits where their growth would be promoted, and we will comment on this work further (\S \ref{prevWork}).

%%%%%%%%%%%%%%%%
\begin{table}[ht]
\caption{Binary system parameters }
\centering
\begin{tabular}{c c c c c c c c}
\hline\hline
System\footnotemark[1] & $M_p$\footnotemark[2]  & $R_p$ & $M_s$ & $R_s$ & $a_b$ & $e_b$ & $a_p$ \\ [0.5ex] % inserts table %heading
\hline 
K34 &  1.05 & 1.16 & 1.02 & 1.09 & 0.23 & 0.52 & 1.09 \\ 
K16 &  0.69 & 0.65 & 0.20 & 0.23 & 0.22 & 0.16 & 0.70 \\ 
K47 &  1.04 & 0.96 & 0.36 & 0.35 & 0.084 & 0.024 & 0.30, 0.99\footnotemark[3] \\ 
K38 &  0.95 & 1.76 & 0.25 & 0.23 & 0.15 & 0.10 & 0.48 \\ 
KIC 4862625 &  1.47 & 1.7 & 0.37 & 0.34 & 0.18 & 0.20 & 0.64 \\ 
K 413 &  0.82 & 0.78 & 0.54 & 0.48 & 0.10 & 0.037 & 0.36 \\ 
K35 &  0.89 & 1.03 & 0.81 & 0.79 & 0.18 & 0.14 & 0.60 \\
KIC 9632895 & 0.93 & 0.83 & 0.19 & 0.2 & 0.18 & 0.05 & 0.79\\ [1ex]
\hline
\end{tabular}
\footnotetext[1]{References: \citet{K47, K34and35, K16, K38, KIC4862625, K413, Welsh14}}
\footnotetext[2]{$M_p$ ($M_s$) is the mass of primary (secondary) star, $R_p$ ($R_s$) is the radius of primary (secondary) star, $a_b$ is the binary semi-major axis,  $e_b$ is the binary eccentricity, $a_p$ is the planetary semi-major axis.}
\footnotetext[3]{This system contains at least two planets.}
\label{table:sysParams}
\end{table}
\par
%%%%%%%%%%%%%%%%

Recently \citet{R13} proposed that the gravitational effect of a massive {\it axisymmetric} protoplanetary disk may strongly suppress eccentricities of circumbinary planetesimals. That happens because disk gravity drives rapid relative precession of planetesimal and binary orbits, reducing the time-averaged non-axisymmetric component of the binary potential, which drives planetesimal eccentricity. However, simulations show that circumbinary disks do not remain axisymmetric and tend to develop eccentricity themselves \citep{Pelupessy13, Meschiari2014}.   Silsbee \& Rafikov (2015; hereafter SR15), developed a formalism to calculate the gravitational effect of an eccentric disk, and applied those results to planet formation in S-type systems.  Rafikov \& Silsbee (2015a; hereafter RS15a), additionally included the effects of gas drag, which damps free eccentricity and produces apsidal alignment of equal-size planetesimals.  The dynamical results of RS15a were applied in Rafikov and Silsbee (2015b; hereafter RS15b) to study collision outcomes in S-type planetary systems.  

In this paper we extend these calculations to P-type systems. In the following discussion of planetesimal dynamics we account for the combined effects of (1) gas drag, (2) gravitational perturbations from the central binary and (3) gravitational perturbations from the eccentric protoplanetary disk, and derive expressions for the encounter velocities of circumbinary planetesimals.  We then use these results to determine which sizes of planetesimals are able to grow at different locations in the disk through mutual collisions.  
\par
In Section \ref{sect:general}, we describe our model system.  In Section \ref{sect:eqs} we write down the disturbing function characterizing the gravitational perturbations from the binary and the disk, and  present the equations governing planetesimal dynamics.  In Sections \ref{sect:exactNoDiskPrecess} - \ref{sect:gen_dyn}, we describe planetesimal dynamics in several different regimes.  Section \ref{sect:CollisionalOutcomes} discusses collision outcomes calculated using the prescription described in Appendix \ref{sect: collOut}.  Section \ref{sect: discussion} critically assesses our underlying assumptions, and explores the outcome of relaxing some of them.  Section \ref{sect:comp} compares our results with those in the literature.  Finally, we summarize our main conclusions about circumbinary planet formation in Section \ref{sect:summary}.

%%%%%%%%%%%%%%%%%%%%%%%%%%%%%%%%%%%%%%%%%%%%%
%%%%%%%%%%%%%%%%%%%%%%%%%%%%%%%%%%%%%%%%%%%%%

\section{General setup}
\label{sect:general}

%%%%%%%%%%%%%%%%%%%%%%%%%%%%%%%%%%%%%%%%%%%%%

Our model system is a close stellar binary consisting of a primary with mass $M_p$ and a secondary with mass $M_s < M_p$ in orbit with semi-major axis $a_b$ and eccentricity $e_b$. Orientation of the binary is given by the apsidal angle $\varpi_b$ with respect to a fixed reference direction. Binary orientation (and $\varpi_b$) may in general vary in time as a result of binary precession; we consider this possibility in Section \ref{sect:binaryPrecession}. For convenience, we define $M_b = M_s + M_p$, and $\mu = M_s/M_b$.  Throughout this paper, unless otherwise noted, we provide numerical estimates for a fiducial system which has the binary parameters of Kepler 16: $M_p = 0.69 M_{\odot}$, $M_s = 0.2 M_{\odot}$, $a_b = 0.22$ AU, $e_b = 0.16$.

Orbiting the barycenter of the binary is a gaseous protoplanetary disk of mass $M_d$. Simulations generally find disks around binaries with $\mu\sim 1/2$ to be tidally truncated on the inside, resulting in the inner cavity relatively devoid of gas \citep{Artymowicz94,Pelupessy13}. According to these calculations the inner edge of the disk is truncated at the inner radius $a_{\rm in}=(1.7-3.3)a_b$, increasing as $e_b$ increases from 0 to 0.7.  Most of the observed stellar orbits in circumbinary systems are on the low end of that eccentricity range (see Table \ref{table:sysParams}), so we assume that the disk is truncated on the inside at $a_{\rm in} = 2 a_b$.  This is somewhat lower than $a_{\rm in} \approx 3 a_b$ favored by \citet{Pelupessy13}.

Our disk model is analogous to that adopted in \citet{SR15}. We assume that circumbinary disk streamlines are ellipses with a common apsidal line and foci at the barycenter of the binary. The orientation of the disk is then defined by a single apsidal angle $\varpi_d$, which may be a function of time (see Section \ref{sect:diskPrecess}) if the disk precesses as a solid body (see a discussion of this assumption in \S \ref{sect:diskdominatedregimedynamics}). We adopt power law scalings for the surface density at periastron $\Sigma_d$ and eccentricity of gas streamlines $e_d$ given by 
\begin{equation}
\label{eq:sigmae}
\Sigma_d(a) = \Sigma_0 \left(\frac{a_0}{a}\right)^p, \quad e_d(a) = e_0 \left(\frac{a_0}{a}\right)^q,
\end{equation}
where $a$ is the semi-major axis of the elliptical fluid trajectory, and $a_0$ is a reference distance, which we take to be 1 AU. These dependencies lead to non-trivial surface density behavior described in \citet{Statler01}; SR15.  

%%%%%%%%%%%%
\begin{table}[ht]
\caption{Fiducial system parameters}
\centering
\begin{tabular}{c c}
\hline\hline
Parameter & Value \\ [0.5ex] % inserts table %heading
\hline 
$M_p$ & $0.69 M_\odot$ \\
$M_s$ & $0.20 M_\odot$ \\
$\mu$ & $0.22$ \\
$a_b$ & $0.22$ AU \\
$e_b$ & 0.16 \\
$\Sigma_0$ &  3,000 g cm$^{-2}$ \\
$e_0$ & 0.024 \\
$p$ & 1.5 \\
$q$ & 1 \\ [1ex]
\hline
\end{tabular}
\label{table:fidParams}
\end{table} 
%%%%%%%%%%%%

For numerical estimates in this work we adopt the following set of disk parameters: $\Sigma_0$ = 3,000 g cm$^{-2}$, $e_0 = 0.024$, and $\Sigma_d$ and $e_d$ slopes $p = 1.5$ and $q = 1$. These and Kepler-16 binary parameters are listed in Table \ref{table:fidParams} for convenience.

Our choice of $p$ corresponds to the MMSN model of \citet{Hayashi81}. This value of $p$ also follows from the decretion disk model of \citet{Pringle91}, assuming no mass to pass through the inner boundary of the disk \citep{R13}.  Our fiducial values of $q$ and $e_0$ are chosen to correspond to the behavior of the forced eccentricity of free particles orbiting in the potential of the binary with eccentricity $e_b$ \citep{MN04}, see Equation \eqref{eq:eforced}. This estimate is just a reasonable zeroth order guess for $e_d(a)$, as eccentricity of the fluid disk is also affected by pressure, viscous forces and disk gravity.

Assuming the mass to be concentrated in the outer part of the disk (i.e. $p < 2$), we can relate $M_d$ to the outer radius of the disk $a_{\rm out}$ and $\Sigma_0$:
\begin{eqnarray}
M_d & = & \frac{2 \pi }{2 - p} \Sigma_0 a_0^pa_{\rm out}^{2-p} 
\label{eq:M_d}\\
%%%%%%%
& \approx & 0.037 M_\odot ~\frac{\Sigma_0}{3,\!000\;{\rm g\;cm^{-2}}} \left(\frac{a_{\rm out}}{75 {\rm AU}}\right)^{0.5},
\end{eqnarray}
where the numerical estimate has been performed for $p = 1.5$.  The outer truncation radius of 75 AU is characteristic of the more massive disks around single stars \citep{Andrews09, Harris12}.

%%%%%%%%%%%%%%%%%%%%%%%%%%%%%%%%%%%%%%%%%%%%%
%%%%%%%%%%%%%%%%%%%%%%%%%%%%%%%%%%%%%%%%%%%%%

\section{Equations of planetesimal dynamics}
\label{sect:eqs}

%%%%%%%%%%%%%%%%%%%%%%%%%%%%%%%%%%%%%%%%%%%%%

We now outline the mathematical framework that we use to describe planetesimal dynamics.  A planetesimal orbit is characterized by the semi-major axis $a_p$ with respect to the binary barycenter, eccentricity $e_p$, and apsidal angle $\varpi_p$; its mean motion is $n_p = \sqrt{GM_b/a_p^3}$. 

Planetesimal motion is affected by both conservative forces --- gravity of the binary and the disk --- and non-conservative gas drag. Analogous to \cite{R13}, SR15, RS15a, we employ secular perturbation theory \citep{MurrayAndDermott,MN04} to determine the effect of the former on planetesimal eccentricity behavior. This approach uses the disturbing function, described next in \S \ref{sect:disturbingfunction}, which accounts for the perturbations to planetesimal motion produced by the gravity of a massive protoplanetary disk and the non-Newtonian gravity of the binary. Our treatment of the effects of gas drag on planetesimal dynamics is described in \S \ref{sect:gas-drag}. All these components are combined in \S \ref{sect:ev-eq}, resulting in a set of general equations (\ref{DDEquations1})-(\ref{DDEquations2}) describing secular planetesimal dynamics in circumbinary planetesimal disks.

%%%%%%%%%%%%%%%%%%%%%%%%%%%%%%%%%%%%%%%%%%%%%

\subsection{Disturbing Function}
\label{sect:disturbingfunction}

%%%%%%%%%%%%%%%%%%%%%%%%%%%%%%%%%%%%%%%%%%%%%

Planetesimal disturbing function $R=R_d+R_b$ consists of contributions due to an eccentric disk $R_d$ and due to the non-Newtonian gravity of the binary $R_b$. 

SR15 derived the disturbing function due to a disk with eccentricity and surface density given by Equation \eqref{eq:sigmae}, and distance-independent apsidal angle $\varpi_d$ as 
\ba
\label{GDF}
R_d &=& n_pa_p^2\left[\frac{A_d}{2} e_p^2 + B_d e_p \cos{(\varpi_p - \varpi_d)}\right],\\
%%%%%
\label{eq:Ad}
A_d & = &  2 \pi \psi_1 \frac{G \Sigma_d(a_p) }{a_p n_p} \\
%%%%%
& \approx & -1.6\times 10^{-3} {\rm yr}^{-1} \nonumber \\
%%%%%
& \times &\frac{\Sigma_0}{3,\!000\; {\rm g\;cm^{-2}}} \frac{\psi_1}{(-0.55)} \left(\frac{0.89M_\odot}{M_b}\right)^{0.5} a_{p,5}^{-1},\nonumber\\
%%%%%
\label{eq:Bd}
B_d & = &  \pi \psi_2 \frac{G \Sigma_d(a_p) e_d(a_p)}{a_p n_p} \\
%%%%%
& \approx &  1.3 \times 10^{-5}{\rm yr}^{-1} \nonumber \\ 
%%%%%
&\times &\frac{\Sigma_0}{3,\!000 \;{\rm g\;cm^{-2}}} \frac{\psi_2}{1.85} \left(\frac{0.89M_\odot}{M_b}\right)^{0.5}  \frac{e_0}{0.024} a_{p,5}^{-1}, \nonumber
\ea
and $a_{p,5}\equiv a_p/(5$AU). Here $\psi_1$ and $\psi_2$ are coefficients of order unity (SR15), which are effectively constants far from the disk edges, for $a_{\rm in}\lesssim a_p\lesssim a_{\rm out}$.  For $p = 1.5$ and $q = 1$, SR15 show that $\psi_1\approx -0.55$ and $\psi_2\approx 1.85$, except near the edges of the disk.  As shown in SR15, $A_d$ and $B_d$ are dominated by local disk properties for our choice of $p$ and $q$, so non-power-law behavior of surface density or eccentricity near the edges of the disk will not greatly affect the values of $\psi_1$ and $\psi_2$. 

According to \citet{MN04} the disturbing function due to the binary has the form similar to Equation \eqref{GDF}:
\begin{eqnarray}
\label{GDFB}
R_b & = & n_pa_p^2\left[\frac{A_b}{2} e_p^2 + B_b e_p \cos{(\varpi_p - \varpi_b)}\right],\\
%%%
\label{AB}
A_b & = & \frac{3}{4} \mu (1-\mu) \frac{n_b^2}{n_p} \left(\frac{a_b}{a_p}\right)^5 \\
& \approx & 1.3\times 10^{-4} {\rm yr}^{-1} 
 \frac{\mu (1- \mu)}{0.17} \left(\frac{a_b}{0.22 {\rm AU}}\right)^2\nonumber \\
& \times & \left(\frac{M_b}{0.89 M_\odot}\right)^{0.5} a_{p,5}^{-3.5},
\nonumber\\
%%%
\label{BB}
 B_b & = & \frac{15}{16} \mu (1-\mu)(1-2\mu) \frac{n_b^2}{n_p} \left(\frac{a_b}{a_p}\right)^6 e_b  \\
& \approx & 6.5\times 10^{-7} {\rm yr}^{-1} \frac{f(\mu)}{0.096}\left(\frac{a_b}{0.22 {\rm AU}}\right)^3 \nonumber \\
& \times & \left(\frac{M_b}{0.89 M_\odot}\right)^{0.5} \frac{e_b}{0.16}a_{p,5}^{-4.5}.
\nonumber
\end{eqnarray}
Here $n_b = \sqrt{GM_b/a_b^3}$ is the mean motion of the binary, and $f(\mu) \equiv \mu(1-\mu)(1-2\mu)$. We use these disturbing functions in Sections \ref{sect:exactNoDiskPrecess} -  \ref{sect:binaryPrecession} to calculate the orbital dynamics of planetesimals in several different dynamical regimes.

In the absence of gas drag and disk gravity, free particles in the binary potential attain forced eccentricity \citep{MN04}
\ba
\label{eq:eforced}
e_{\rm forced} &=& \frac{B_b}{A_b} = \frac{5}{4} (1-2 \mu) \frac{a_b}{a_p} e_b\\
%%%%
&\approx & 0.024~\frac{(1- 2\mu)}{0.56} \frac{e_b}{0.16}~
\frac{a_b}{0.22 {\rm AU}}~\frac{1 {\rm AU}}{a_p},
\nonumber
\ea
which provides a useful reference value, e.g. for our choice of $e_0$.  

Introducing planetesimal eccentricity vector ${\bf e}_p=(k_p,h_p)=e_p(\cos\varpi_p,\sin\varpi_p)$ we can then write down the full disturbing function as 
\begin{eqnarray}
R = n_pa_p^2 && \left[\frac{A}{2} (h_p^2 + k_p^2) + B_dk_p\cos\varpi_d \right.
\label{eq:full_dist_func}\\
%%%%%%%%%
&&+ \left. B_dh_p\sin\varpi_d+ B_bk_p\cos\varpi_b + B_bh_p\sin\varpi_b\right],
\nonumber
\end{eqnarray}
where $\varpi_d=\varpi_d(t)$, $\varpi_b=\varpi_b(t)$. 

\begin{figure}
\centering
\includegraphics[scale=.48]{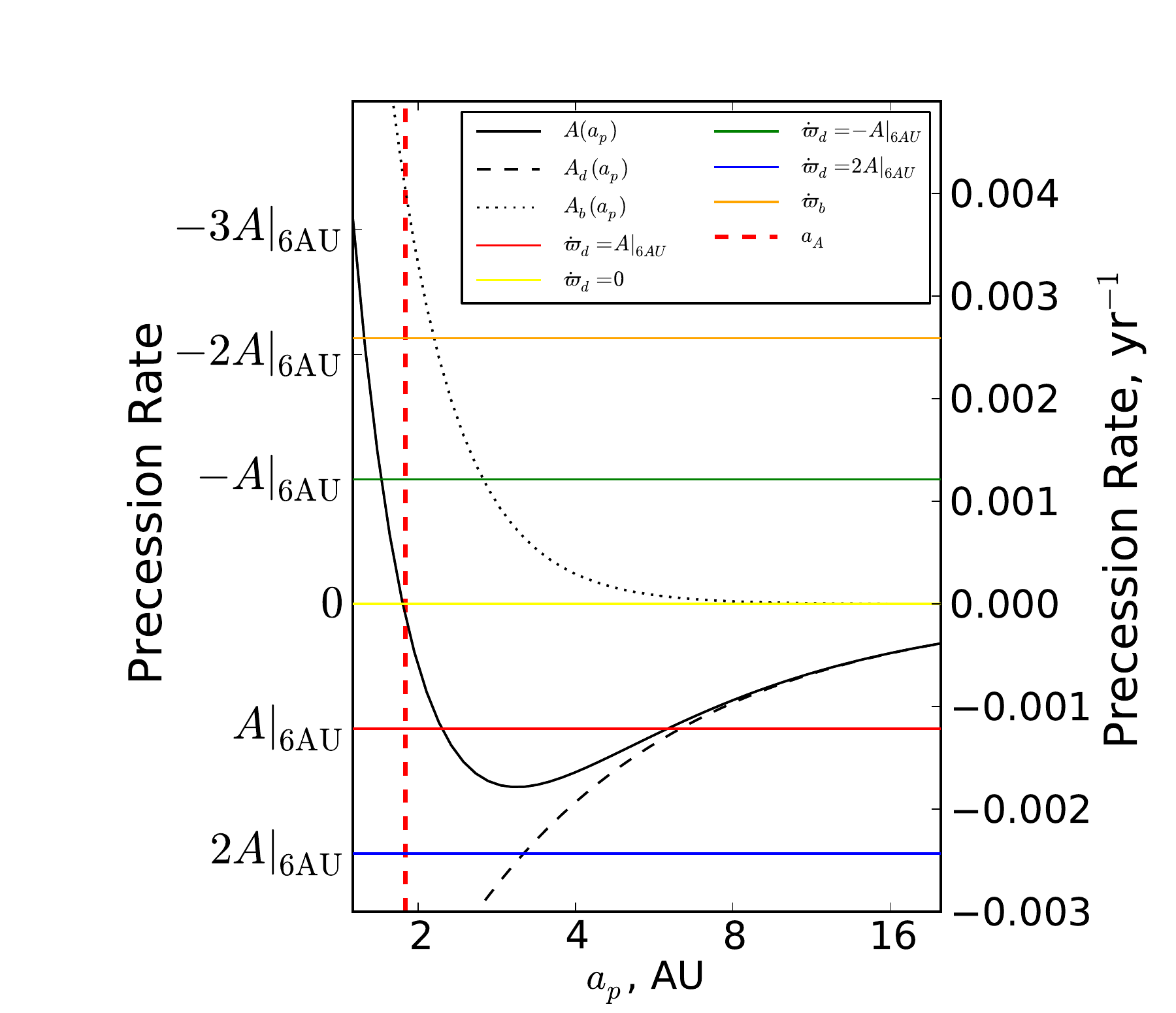}
\caption{Planetesimal precession rate $A=A_b+A_d$ due to both binary and disk gravity (black curve) as a function of $a_p$, calculated assuming the fiducial system parameters in Table \ref{table:fidParams}. Dotted and dashed curves represent $A_d(a_p)$ and $A_b(a_p)$. Several reference values corresponding to disk and binary precession rates considered in \S \ref{sect:diskPrecess} and \S \ref{sect:generalbinpre} are also shown (expressed in units of $A|_{\rm 6AU}<0$ --- planetesimal precession rate at 6 AU).  The vertical dashed line marks $a_p=a_A$, see equation (\ref{eq:aequality}).  For our fiducial binary + disk model (see Table \ref{table:fidParams}) $A|_{6AU} = -1.2\times 10^{-3} \, {\rm yr^{-1}}$, $\dot \varpi_b = 2.6 \times 10^{-3} \, {\rm yr^{-1}}$.}
\label{resonances}
\vspace{-.05cm}
\end{figure}

Here $A = A_d + A_b$ is a precession rate of planetesimal free eccentricity due to the axisymmetric components of both the disk and binary gravity. Its behavior as a function of distance from the binary is shown in Figure \ref{resonances}, together with separate curves for $A_d(a_p)$ and $A_b(a_p)$. The fact that $A$ goes through zero at some semi-major axis $a_A$ has very important implications for planetesimal dynamics, see \S \ref{sect:bin-disk-transition}.

%%%%%%%%%%%%%%
%%%%%%%%%%%%%%

The other terms in equation (\ref{eq:full_dist_func}) that depend on the planetesimal orientation (i.e. $\varpi_p$) describe excitation of planetesimal eccentricity by the torques produced by the non-axisymmetric components of the disk and binary potentials. We discuss the relative role of different contributions to $R$ next.

%%%%%%%%%%%%%%%%%%%%%%%%%%%%%%%%%%%%%%%%%%%%%

\subsubsection{Transition Between Binary-Dominated and Disk-Dominated Regimes}
\label{sect:bin-disk-transition}

%%%%%%%%%%%%%%%%%%%%%%%%%%%%%%%%%%%%%%%%%%%%%

Because $A_b$ and $B_b$ fall off more rapidly with semi-major axis than their disk-related counterparts $A_d$ and $B_b$, see Equations (\ref{eq:Ad})-(\ref{eq:Bd}) and (\ref{AB})-(\ref{BB}), we find that outside a certain radius the disturbing function $R$ should be dominated by the disk gravity. This situation is analogous to the so-called DD regime discussed in SR15, in which both the axisymmetric and the non-axisymmetric components of the disturbing function are dominated by the disk.  

In the opposite limit, very close to the binary, the gravitational perturbations are dominated by the binary system. This is analogous to the so-called BB regime of SR15 in which both the axisymmetric and the non-axisymmetric components of the disturbing function are dominated by the binary. At the same time, the disk still cannot be ignored because of the gas drag.   

Equations (\ref{eq:Ad}) and (\ref{AB}) predict that planetesimal precession (i.e. $A$) switches from being binary- to disk-dominated at a characteristic semi-major axis $a_A$, where $|A_d|=|A_b|$:
\ba
\label{eq:aequality}
a_A &=& \left[\frac{3 \mu(1-\mu)}{8 \pi |\psi_1|}\frac{M_ba_b^2}{\Sigma_0 a_0^p}\right]^{1/(4-p)}  
%%%%%
\\
& \approx & 1.9 {\rm AU} \left[\left(\frac{a_b}{0.22{\rm AU}}\right)^2 \frac{M_b}{0.89 M_\odot} 
\right.
\nonumber\\
%%%%%
&& \left.~~~~~~~~\times \frac{3,\!000{\rm g cm^{-2}}}{\Sigma_0}  
\left(\frac{\mu (1- \mu)}{0.17}\right) \right]^{2/5},  \nonumber
\ea
where the numerical estimates have been performed for our fiducial system.  This transition occurs via a secular resonance where $A=0$, emerging because disk and binary drive planetesimal precession in opposite directions, see equations (\ref{eq:Ad}) and (\ref{AB}). 

On the other hand, planetesimal eccentricity excitation switches from being binary- to disk-dominated at the different characteristic distance $a_B$, where $|B_d|=|B_b|$:
\ba
\label{eq:bequality}
a_B &=& \left[\frac{15 f(\mu)}{16 \pi |\psi_2|} \frac{M_b a_b^3}{\Sigma_0 a_0^{p+q}} \frac{e_b}{e_0}\right]^{1/(5-p-q)} \\ 
%%%%%%%%
& \approx & 1.5 {\rm AU} \left[\frac{e_b}{0.16} \frac{0.024}{e_0} \frac{M_b}{0.89 M_\odot} \left(\frac{a_b}{0.22 {\rm AU}}\right)^3 \right.
\nonumber\\
%%%%%
&& \left.~~~~~~~~\times\frac{3000\;{\rm g\;cm^{-2}}}{\Sigma_0}\frac{f(\mu)}{0.096}\right]^{2/5}. \nonumber 
\ea
For a single component of the disturbing function (axisymmetric or non-axisymmetric), the region of transition between binary-domination and disk-domination is quite narrow, since both $A_d/A_b$ and $B_d/B_b$ are rising fast with $a_p$ (as $a_p^{5/2}$). 

%%%%%%%%%%%%%%
\begin{figure}
\centering
\includegraphics[scale = .4]{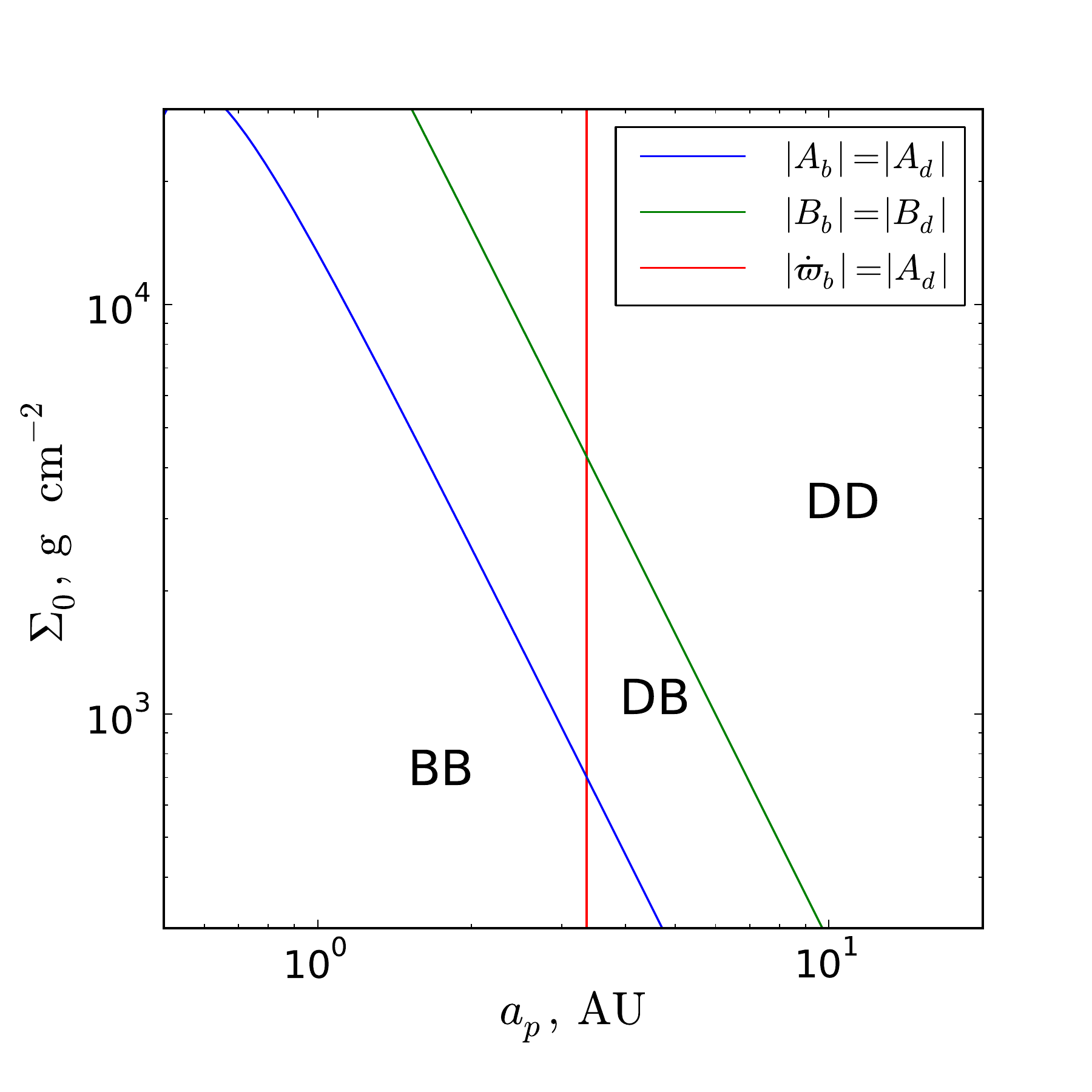}
\caption{Three different dynamical regimes in the space of disk density and semi-major axis. Calculation is done for our fiducial system parameters ($M_p = 0.69 M_{\odot}$, $M_s = 0.2 M_{\odot}$, $a_b = 0.22$ AU,  $e_b = 0.16$,  $p = 1.5$, and $q = 1$), but with lowered disk eccentcity at 1 AU $e_0=2.4 \times 10^{-3}$, which is 10\% of the forced eccentricity (\ref{eq:eforced}) to broaden the DB regime. The vertical red line shows the location at which binary precession rate equals to the disk-driven planetesimal precession, the significance of which is discussed in \S \ref{sect:axisymmetricdiskdynamics}.}
\label{phaseSpaceFigure}
\vspace{-.05cm}
\end{figure}
%%%%%%%%%%%%%%

If $a_A$ is widely separated from $a_B$, there will also be a region of the disk where excitation is dominated by the disk and precession by the binary, or vice-versa.  In particular, in Section \ref{sect:axisymmetricdiskdynamics}, we consider the limit of a low eccentricity disk ($e_0\to 0$) in which $a_A$ can be substantially less than $a_B$, see Equations \eqref{eq:aequality}, and \eqref{eq:bequality}.  Planetesimal dynamics in the intermediate region $a_A \lesssim a_p \lesssim a_B$ are then analogous to the DB regime of SR15. In the opposite limit of a high-eccentricity disk, $e_0=e_d(1~\mbox{AU})\gtrsim 0.1$ (which is probably not very realistic), there would exist a region $a_B \lesssim a_p \lesssim a_A$, in which the eccentricity excitation is dominated by the disk, while precession is controlled mainly by the binary --- analogous to the BD regime of SR15.  

The locations of these regimes in the space of disk density and eccentricity are illustrated in Figure \ref{phaseSpaceFigure}.  The width of the DB regime depends on the degree to which the eccentricity of the disk falls below the free particle eccentricity in the binary potential. Figure \ref{phaseSpaceFigure} also includes the line where $|\dot \varpi_b| = |A_d|$.  This gives roughly the radius outside of which excitation due to the binary is substantially reduced by its precession (see Section \ref{sect:binaryPrecession}).

%%%%%%%%%%%%%%%%%%%%%%%%%%%%%%%%%%%%%%%%%%%%%

\subsection{Gas drag}
\label{sect:gas-drag}

%%%%%%%%%%%%%%%%%%%%%%%%%%%%%%%%%%%%%%%%%%%%%

Following RS15a, we consider gas drag to damp planetesimal eccentricity as
\ba
\frac{d{\bf e}_p}{dt}=-\frac{{\bf e}_p-{\bf e}_d}{\tau_d},
\label{eq:drag}
\ea
where ${\bf e}_d=(k_d,h_d)=e_d(\cos\varpi_d,\sin\varpi_d)$ is the eccentricity vector of the local disk fluid element, and $\tau_d$ is the eccentricity damping time. This implies that gas drag drives planetesimal orbits towards full alignment with the gas trajectories on a characteristic timescale $\tau_d$. 

This timescale is not independent of ${\bf e}_p$ in general. For planetesimals with radii $d_p$ on the order of km, a quadratic drag law ${\bf F}=-(C_D/2)\pi d_p^2\rho_d v_r{\bf v}_r$ is appropriate \citep{Weidenschilling77}, where $C_D\approx 0.5$ is the drag coefficient, $\rho_d$ is the local gas density and ${\bf v}_r$ is the relative planetesimal-gas speed. In this case $\tau_d\propto v_r^{-1}$, and RS15a had $\tau_d\propto e_r^{-1}$, where ${\bf e}_r={\bf e}_p-{\bf e}_d$ is the relative eccentricity between the object and the gas.

In this work we have chosen to also account for the fact that gas orbits the binary {\it more slowly} than a planetesimal because of the radial pressure support in a gaseous disk. This gives rise to additional (predominantly azimuthal) irreducible relative velocity between the gas and planetesimal $\Delta v_{\phi}=\eta v_K$, where $v_K$ is the Keplerian speed and the explicit expression for $\eta\ll 1$ is given by eq. (13) of RS15b. This velocity differential does not vanish even when fluid and planetesimal orbits coincide and ${\bf e}_p={\bf e}_d$.  To describe this effect, we introduce fiducial eccentricity 
\ba
e_\phi &=& \frac{\pi }{2 E(\sqrt{3}/2)}\frac{\Delta v_{\phi}}{v_K},
\label{eq:e_phi}\\
%%%%%
&\approx & 0.0038 \frac{M_\odot}{M_b} a_{p, 5}^{1/2}.
\nonumber
\ea
The numerical estimate uses the prescription for the scale height $h$ given in RS15b (their eq. (14)), adapted for a disk temperature\footnote{Stars in P-type binaries typically have lower masses and lower luminosities that the members of S-type binaries studied in SR15b.} of 200 K at 1 AU:  
\begin{equation}
\label{eqhoverr}
\frac{h}{a_p} = 0.028 \sqrt{\frac{M_{\odot}}{M_b}} \left(\frac{a_p}{\rm AU}\right)^{1/4}.
\end{equation}
\par
We generalize the expression for the characteristic damping time $\tau_d$ from RS15a as follows:
\ba
\tau_d &=& \frac{2^{5/2}\pi^{3/2}}{3 C_D {\rm E}(\sqrt{3}/2)}n_p^{-1} \frac{\rho_p d_p}{\Sigma_d}\frac{h}{a_p} \left(e_r^2 + e_\phi^2\right)^{-1/2}
\label{taud}\\
& \approx &1.6 \times 10^5 \, {\rm yr}  \frac{\rho_p}{3 \, {\rm g\, cm}^{-3}}  \frac{d_p}{\rm km} \nonumber \\
& \times &
\frac{0.89 M_\odot}{M_b} \frac{3,\!000 \, {\rm g\, cm}^{-2}}{\Sigma_0}  \frac{0.01}{\sqrt{e_r^2 + e_\phi^2}} a_{p,5}^{13/4}.
%%%%%
\nonumber
\ea
Here, E$(\sqrt{3}/2)\approx 1.21$ is a complete elliptic integral, $\rho_p$ is the planetesimal bulk density, and we have assumed $p = 1.5$ in the estimate.

Note that the numerical coefficient in equation (\ref{taud}) is different from the analogous expression in RS15a because they adopted a rough estimate $\rho_d = \Sigma_d/h$, whereas here we use $\rho_g = \Sigma_d/\sqrt{2 \pi}h$, appropriate for an isothermal disk. We have picked the coefficient in the definition of $e_\phi$ to give the correct damping time (in agreement with \citet{Adachi76}) in the limit that $e_r \ll e_\phi$.  

Equation \eqref{taud} is only approximate for $e_\phi \sim e_r$. However, it does capture the reduction of $\tau_d$ for particles with low values of $e_r$ due to the irreducible velocity differential $\Delta v_\phi$ caused by pressure support. It is thus an improvement over the approximation used in SR15a. Low values of $e_r\lesssim e_\phi$ are more typical at several AU around P-type binaries than in S-type systems.

%%%%%%%%%%%%%%%%%%%%%%%%%%%%%%%%%%%%%%%%%%%%%

\subsection{General evolution equations}
\label{sect:ev-eq}

%%%%%%%%%%%%%%%%%%%%%%%%%%%%%%%%%%%%%%%%%%%%%

We now combine the results of \S \ref{sect:bin-disk-transition} and \ref{sect:gas-drag}. We use the standard Lagrange equations \citep{MurrayAndDermott,R13} to relate $dk_p/dt$ and  
$dh_p/dt$ to the disturbing function (\ref{eq:full_dist_func}). Adding the contributions due to gas drag given by equation (\ref{eq:drag}), we find, to lowest order in eccentricity, the following set of evolution equation for planetesimal eccentricity ${\bf e}_p$:

\ba
\frac{dk_p}{dt} &=& -Ah_p-B_d\sin\varpi_d(t)-B_b\sin\varpi_b(t) 
\nonumber\\
%%%%%%%
&-& \frac{k_p - k_d}{\tau_d}, 
\label{DDEquations1}\\
%%%%%%%
\frac{dh_p}{dt} &=& Ak_p + B_d\cos\varpi_d(t)+ B_b\cos\varpi_b(t) 
\nonumber\\
%%%%%%%
&-& \frac{h_p - h_d}{\tau_d}.
\label{DDEquations2}
\ea
As before, $\varpi_d$ and $\varpi_d$ are in general functions of time. 

These master equations provide a basis for subsequent analysis of planetesimal dynamics in Sections \ref{sect:exactNoDiskPrecess} -  \ref{sect:binaryPrecession}.

%%%%%%%%%%%%%%%%%%%%%%%%%%%%%%%%%%%%%%%%%%%%%
 
\subsection{General Remarks on Planetesimal Dynamics}
\label{sect:dynamicsgen}

%%%%%%%%%%%%%%%%%%%%%%%%%%%%%%%%%%%%%%%%%%%%%

Before embarking on a detailed discussion of planetesimal dynamics in the following sections, we outline some general features of planetesimal eccentricity evolution described by equations (\ref{DDEquations1})-(\ref{DDEquations2}).

First, in the case of quadratic drag these equations do not in general admit an analytical solution for arbitrary time dependence of $\varpi_d(t)$ and $\varpi_b(t)$. At the same time, there are several important limits where analytical treatment is possible, and these situations are covered in \S \ref{sect:exactNoDiskPrecess}, \ref{sect:diskdominatedregimedynamics}, \ref{sect:axisymmetricdiskdynamics}. These solutions allow us to gain important insights into how the binary or disk precession may affect planetesimal dynamics, which remain valid in more complicated setups (\S \ref{sect:gen_dyn}). In addition, some features of the general planetesimal dynamics with both $\varpi_d$ and $\varpi_b$ varying in time can be gleaned by considering a simpler case of linear gas drag, covered in Appendix \ref{sect:lindrag} and discussed in \S \ref{sect:gen_dyn}.

Second, we find quite generally that any free eccentricity describing the initial conditions for planetesimal evolution damps away on a characteristic time $\sim \tau_d$. As a result, planetesimal eccentricity ${\bf e}_p$ inevitably converges to its forced value, which is determined by many factors, see \S \ref{sect:exactNoDiskPrecess}. This feature of the evolution has been previously pointed out in \citet{Beauge10} and RS15a, and implies that after the initial transient lasting for $\sim \tau_d$ planetesimals lose memory of their initial conditions. 

In particular, collisions between planetesimals, which perturb them away from the equilibrium eccentricities, may be considered as a minor effect for the dynamics as long as they are infrequent enough, i.e. the mean time between them is longer than $\tau_d$.  We provide a discussion of this approximation in Section \ref{sect: limits of apsidally aligned}. Convergence of ${\bf e}_p$ to a certain fixed state greatly simplifies our analysis as we see in the following sections.

Circumbinary systems exhibit a wide range of planetesimal dynamical behavior throughout the disk. We now describe them under different assumptions about the disk and binary precession. We start by considering a simple case where we ignore the effects of binary precession. Although the latter is likely very important in reality, ignoring it at first allows us to better illustrate certain aspcts of planetesimal dynamics. Thus, we consider the case of non-precessing disk and binary in \S \ref{sect:exactNoDiskPrecess}. We then include the possibility of the disk precession in \S \ref{sect:diskdominatedregimedynamics}. Finally, in \S \ref{sect:binaryPrecession} and \S \ref{sect:gen_dyn} we consider a possibility of the binary precession.

%%%%%%%%%%%%%%%%%%%%%%%%%%%%%%%%%%%%%%%%%%%%%
%%%%%%%%%%%%%%%%%%%%%%%%%%%%%%%%%%%%%%%%%%%%%
 
\section{Planetesimal Dynamics in the Absence of Disk and Binary Precession}
\label{sect:exactNoDiskPrecess}

%%%%%%%%%%%%%%%%%%%%%%%%%%%%%%%%%%%%%%%%%%%%%

If we ignore precession of both the disk and the binary, i.e. take $\varpi_d(t)$ and $\varpi_b(t)$ to be constant in time, we can easily solve equations (\ref{DDEquations1})-(\ref{DDEquations2}). This approximation should be valid for small planetesimals, which have stopping times shorter than the two precession timescales, $\tau_d\lesssim \dot\varpi_d, \dot \varpi_b$. In this case the eccentricity vector ${\bf e}_p$ should rapidly adjust to the fixed values corresponding by the {\it instantaneous} values of $\varpi_d$ and $\varpi_b$. 

Solutions for planetesimal dynamics in this limit accounting for both the disk and the binary gravity have been previously derived in RS15a, but for S-type binaries. They remain fully valid in the circumbinary case as well, as long as $A_b$ and $B_b$ are understood as being represented by our Equations \eqref{eq:Ad} and \eqref{eq:Bd}. The forced eccentricity to which ${\bf e}_p$ converges after the initial period of damping the free eccentricity, is given by a sum of forced terms due to the binary ${\bf e}_{f,b}$ and to the disk ${\bf e}_{f,d}$:
\begin{eqnarray}
{\bf e}_p & = &  \left\{
\begin{array}{l}
k_p\\
h_p
\end{array}
\right\}={\bf e}_{f,b}+{\bf e}_{f,d},
\label{eq:vect_sum}
\\
%%%%%
{\bf e}_{f,b} & = &
\left[\frac{\tau_d^2 B_b^2}{1+(A\tau_d)^2}\right]^{1/2}
\left\{
\begin{array}{l}
\cos\left(\varpi_b+\phi_b\right)\\
\sin\left(\varpi_b+\phi_b\right)
\end{array}
\right\},
\label{eq:e_e_b}\\
%%%%%
{\bf e}_{f,d} & = & 
\left[\frac{e_d^2+\tau_d^2 B_d^2}{1+(A\tau_d)^2}\right]^{1/2}
\left\{
\begin{array}{l}
\cos\left(\varpi_d+\phi_d\right)\\
\sin\left(\varpi_d+\phi_d\right)
\end{array}
\right\},
\label{eq:e_f_d}
\end{eqnarray}
where $\phi_d$ and $\phi_b$ are phase angles given by 
\begin{equation}
\cos\phi_d=\frac{e_d- A B_d\tau_d^2}
{\left(e_d^2+\tau_d^2 B_d^2\right)^{1/2}
\left[1+(A\tau_d)^2
\right]^{1/2}},
\label{eq:phid_nonprec}
\end{equation}
and 
\begin{equation}
\cos\phi_b=\frac{- A\tau_d}{\sqrt{1+(A\tau_d)^2}}.
\label{eq:phib_nonprec}
\end{equation}
One can see that in general ${\bf e}_p$ is misaligned with both the disk and the binary as it is parallel to neither ${\bf e}_d$ nor ${\bf e}_b=e_b(\cos\varpi_b,\sin\varpi_b)$. It is also clear that ${\bf e}_p$ does not vary in time when $\varpi_d$ and $\varpi_b$ are fixed.

%%%%%%%%%%%%%%%%%%%%%%%%%%%%%%%%%%%%%%%%%%%%%

\subsection{Relative Planetesimal-Gas Eccentricity}

Solution \eqref{alignedSolution} implies that the relative planetesimal-gas eccentricity $e_r=|{\bf e}_p-{\bf e}_d|$ is given by (RS15a)
\begin{equation}
\label{erAligned}
e_r = e_c \frac{A\tau_d}{\sqrt{1+\left(A\tau_d\right)^2}},
\end{equation}
where the characteristic eccentricity $e_c$ is 
\ba
e_c = |A|^{-1} && \left[(Ae_d + B_d)^2 + B_b^2 \right.
\nonumber\\
%%%%%%
&& \left.+ 2 \cos{(\varpi_d - \varpi_b)} B_b(Ae_d + B_d)\right]^{1/2}.
\label{eq:ecRS15a}
\ea

We find it convenient to define a characteristic size $d_c$ for which $A\tau_d=1$ when $\sqrt{e_r^2+e_\phi^2}$ is replaced with $e_c$ in Equation \eqref{taud}, i.e. eccentricity damping time is of order the orbit precession time scale.  Equation (\ref{taud}) implies that 
\begin{align}
\label{eq:dc}
& d_c = \frac{3 C_D E(\sqrt{3}/2)}{2^{5/2} \pi^{3/2}} \frac{\Sigma_d}{\rho_p}\frac{n_p}{|A|}\frac{a_p}{h} e_c\nonumber\\
& = 1.3{\rm m} \left(\frac{M_b}{0.89 M_\odot}\right)^{1.5}\frac{3\;{\rm g \; cm^{-3}}}{\rho_p} a_{p,5}^{-13/4}.
\end{align}
Because the numerical estimate is made at 5 AU, far outside of $a_A$ and $a_B$ for typical disk parameters, we have assumed for simplicity that the contribution of the binary to the disturbing function is negligible, i.e. $A\approx A_d$, $B_b \approx 0$. In this regime (far from the star) both $e_c$ and $d_c$ are independent of disk mass $M_d$ (or $\Sigma_0$). This can be seen from Equations \eqref{eq:ecRS15a} and \eqref{eq:dc}, and the fact that $A_d$, $B_d$ and $\Sigma_d$ are all proportional to $M_d$. 

Using our newly defined $d_c$, we may rewrite equation (\ref{taud}) as $A\tau_d = (d_p/d_c) \left(e_c/\sqrt{e_r^2 + e_\phi^2}\right)$. Combining this result with Equation \eqref{er} we can solve for $A\tau_d$ and $e_r$:
\ba
\label{eq:atd}
A_d\tau_d & = & 2^{-1/2}\left[\frac{1-K + L}{(d_c/d_p)^2 + K}\right]^{1/2},
\\
%%%%%%%%%%
\label{eq:erInTermsOfDp}
e_r & = & e_c \left[\frac{1-K + L}{1+K + 2 (d_c/d_p)^2 + L}\right]^{1/2},
\ea
where 
\ba
K & \equiv & (e_\phi/e_c)^2(d_c/d_p)^2,
\nonumber\\
L &\equiv & \left[(K+1)^2 + 4(d_c/d_p)^2\right]^{1/2}.
\nonumber
\ea 

For most of the paper, we will be considering situations in which $e_c \gg e_\phi$.  In this limit, $K \rightarrow 0$, $L \rightarrow 2 d_c/d_p$, and planetesimal dynamics are determined purely by the values of $d_c$ and $e_c$, so that our results reduce to those of RS15a, with $\tau_d$ longer by a factor of $\sqrt{2 \pi}$ (to compensate for the different definition of $\rho_d$).

%%%%%%%%%%%%%%%%%%%%%%%%%%%%%%%%%%%%%%%%%%%%%

\subsection{Solution far from the Binary}
\label{sect:np}

A useful limit of planetesimal dynamics without disk or binary precession is obtained when we are justified in neglecting the binary perturbation.  This regime is naturally realized far from the binary.  This is equivalent to the DD regime of SR15 in which disk gravity dominates dynamics, so $|B_d| \gg |B_b|$ and  $A\approx A_d$. 

It is easy to show in this case that except near the edge of the disk (where edge effects change the values of $\psi_1$ and $\psi_2$), $e_c$ becomes a constant multiple of the local disk eccentricity $e_d$. Indeed, if the precession of the disk is ignored, then Equation \eqref{eq:ecRS15a} gives us that $e_c = |B_d/A_d + e_d|$.  Using Equations \eqref{eq:Ad} and \eqref{eq:Bd}, this can be rewritten as
\begin{equation}
\label{ddec}
e_c = e_d\left|\frac{\psi_2}{2\psi_1} + 1 \right| \approx 0.65 e_d,
\end{equation}
as for $p = 3/2$, $q = 1$, we have $\psi_2 \approx 1.82$ and  $\psi_1 \approx -0.55$. In this regime, planetesimal survival is easier at larger semi-major axes simply because larger $a_p$ means lower $e_d(a_p)$ for $q > 0$.

%%%%%%%%%%%%%%%%%%%%%%%%%%%%%%%%%%%%%%%%%%%%%

\subsection{Radial Behavior of $e_r$}
\label{sect:fixed_ep}

\begin{figure}
\centering
\includegraphics[scale = .5]{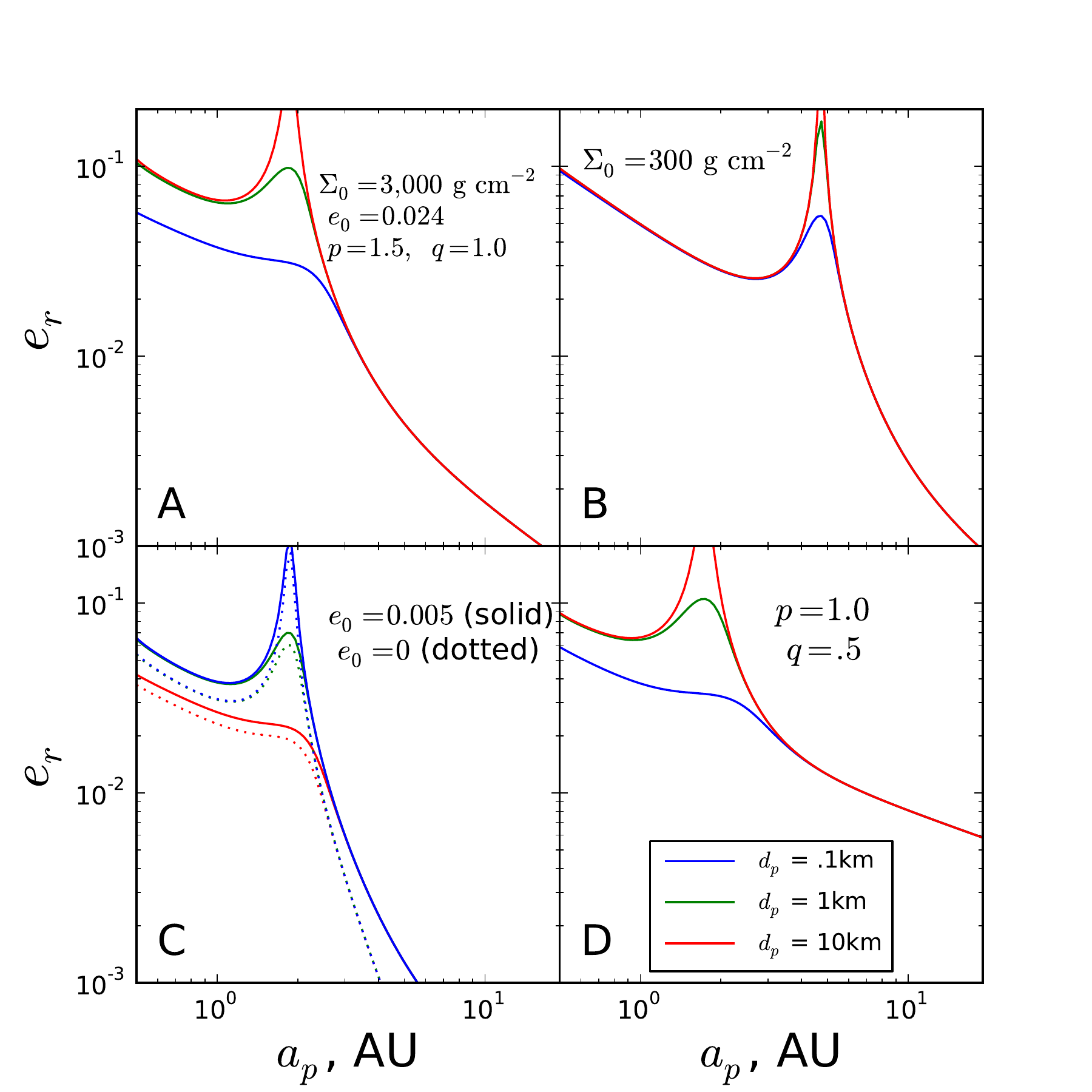}
\caption{Radial dependence of $e_r$ (see Equation (\ref{erAligned})) in a system with no binary or disk precession, but including gravity of both the binary and the disk, for 4 different disk models. Calculations assume the binary parameters of Kepler 16 with fiducial disk parameters unless otherwise labelled; $\varpi_d=\varpi_b$ is assumed. Characteristic $e_c$ is determined by Equation (\ref{eq:ecRS15a}). Different colors correspond to different planetesimal sizes, as indicated in panel D.}
\label{erRun}
\vspace{-.05cm}
\end{figure}

The behavior of the relative planetesimal-gas eccentricity given by equations \eqref{erAligned} and \eqref{eq:erInTermsOfDp} is illustrated in Figure \ref{erRun}, which shows $e_r$ as a function of orbital distance $a_p$ for different disk models. Unless otherwise noted, all panels assume the Kepler 16 binary parameters with our fiducial disk parameters as displayed in Panel A,  and apsidal alignment of the binary and the disk, i.e. $\varpi_d=\varpi_b$. 

The most pronounced feature seen in all panels is the secular resonance, where $A = 0$, see Equation \eqref{eq:ecRS15a}. It inevitably appears in the non-precessing disks at $a_A$ because $A$ changes sign there, as a result of $A_b$ and $A_d$ having different signs, see Figure \ref{resonances}. The resonance location is different in panel B because of the lowered $\Sigma_0$ in this particular disk model, which pushes $a_A$ out to 4.8 AU, compared to $a_A\approx 1.9$ AU in all other panels. Note that although formally $e_c\to \infty$ at the resonance, $e_r$ stays finite there. Moreover, the increase of $e_r$ at the resonance is often not very pronounced: unless $d_p > d_c$, the resonance has little effect on relative eccentricities because the planetesimals are closely coupled to the gas (see Equation \eqref{eq:erInTermsOfDp}).  The existence of the secular resonance previously suggested in the drag-free environment by R13 was confirmed numerically by \citet{Meschiari2014}. In his simulations disk gravity excites the eccentricities of $d_p=5$ km planetesimals in a radially narrow disk region around $a_p\approx 2.5$ AU by a factor of $\sim 10$ compared to $e_p$ at that radius in the absence of disk gravity.

We see that lowering the disk eccentricity (panel C) causes $e_r$ to be dramatically lower outside of the secular resonance. This is because far from the star, disk gravity dominates, and we reach the limit of Section \ref{sect:np}, where $e_c$ is proportional to $e_d$. On the other hand, reduction of $e_0$ has little effect near the star, where excitation is dominated by the binary. This is because very close to the star, terms $A_d$ and $B_d$ are irrelevant in Equation \eqref{eq:ecRS15a}. At the same time, disk eccentricity still enters through $e_d$, which is not negligible (see Equation \eqref{eq:ecRS15a}), so some variation of $e_p$ is still present near the star as $e_0$ is varied.  

Finally, panel D shows a shallower drop-off of planetesimal eccentricity at large $a_p$ simply because we adopted a model with a different (lower) value of $q$, and $e_p$ is in the disk-dominated regime outside the secular resonance.

%%%%%%%%%%%%%%%%%%%%%%%%%%%%%%%%%%%%%%%%%%%%%
%%%%%%%%%%%%%%%%%%%%%%%%%%%%%%%%%%%%%%%%%%%%%

\section{Dynamics with Disk-Dominated Excitation and Disk precession} 
\label{sect:diskdominatedregimedynamics}

%%%%%%%%%%%%%%%%%%%%%%%%%%%%%%%%%%%%%%%%%%%%%

We will find later in this work (\S \ref{sect:pgdder}) that it is generally most promising to form planets outside of a few AU, relatively far from the binary. At large $a_p\gtrsim a_A$ planetesimal eccentricity {\it excitation} is dominated solely by the disk gravity, even though binary gravity may still affect planetesimal {\it precession}, see \S \ref{sect:bin-disk-transition}.

For that reason we will now explore the limit in which the binary plays a negligible role in the planetesimal eccentricity excitation, i.e. $|B_b|\lesssim |B_d|$. At the same time, the binary is still allowed to contribute significantly to planetesimal precession, i.e. $A=A_b+A_d$ in general. Thus, the current limit is analogous to the DD and DB regimes of SR15, with the addition of gas drag and disk precession. 

Here we also include a (realistic) possibility of the disk precession at some constant rate $\dot \varpi_d$. For the purposes of deriving an analytical solution, this rate must be independent of the semi-major axis, because the derivation of $R_d$ in SR15 assumes fluid trajectories to be apsidally aligned at all $a_p$.  In practice this assumption is likely to break far from the binary. However, this is not a problem as the disk eccentricity is going to be very low there anyway \citep{Pelupessy13, Meschiari2014}.  

Setting $\varpi_d = \dot \varpi_d t$ and $B_b\to 0$ in Equations (\ref{DDEquations1})-(\ref{DDEquations2}) we find that, as shown in RS15a, these equations admit a periodic solution in the form 
\ba
{\bf e}_p =  
\left[\frac{e_d^2+\tau_d^2 B_d^2}{1+(A - \dot \varpi_d)^2\tau_d^2}\right]^{1/2}
\left\{
\begin{array}{l}
\cos\left(\dot \varpi_d t+\phi\right)\\
\sin\left(\dot \varpi_d t+\phi\right)
\end{array}
\right\},
\label{alignedSolution}
\ea
where the phase shift $\phi$ with respect to the disk apsidal line is still given by equation (\ref{eq:phid_nonprec}) but with $A$ replaced by $A - \dot \varpi_d$.  In this solution, the forced eccentricity vector ${\bf e}_p$ rotates at the rate $\dot\varpi_d$ and is fixed in the frame precessing with the disk.

Solution \eqref{alignedSolution} implies that the relative planetesimal-gas eccentricity is given by 
\begin{equation}
\label{er}
e_r = e_c \frac{(A - \dot \varpi_d) \tau_d}{\sqrt{1+\left[(A - \dot \varpi_d) \tau_d\right]^2}},
\end{equation}
where the characteristic eccentricity $e_c$ is now
\begin{equation}
\label{eq:ec}
e_c = \left|\frac{B_d}{A - \dot \varpi_d} + e_d\right|.
\end{equation}
Also, equations (\ref{eq:dc})-(\ref{eq:erInTermsOfDp}) still hold with the provision that $A$ is replaced with $A - \dot \varpi_d$ everywhere.

%%%%%%%%%%%%%%%%%%%%%%%%%%%%%%%%%%%%%%%%%%%%%

\subsection{Effects of Disk Precession}
\label{sect:diskPrecess}

%%%%%%%%%%%%%%%%%%%%%%%%%%%%%%%%%%%%%%%%%%%%%

We now explore the effect of disk precession on the behavior of planetesimal eccentricity. We examine the run of $e_r$ in the disk for different disk models in Figure \ref{DiskDominatedErRun}, assuming the binary parameters of Kepler 16. All disks have $p = 1.5$, $q = 1$, and $\Sigma_0 = 3,\!000$ g cm$^{-2}$. Although we show the behavior of $e_r$ starting at 1 AU, our disk-dominated excitation assumption is valid only for $a_p\gtrsim a_B$ (the latter is shown by the green vertical line), which may be a problem for low $e_0$ (left panels in this figure) and has to be kept in mind. Because of that this figure would be accurate for all $a_p$ only if the binary eccentricity were zero.

As a fiducial value of $\dot \varpi_d$ we take the planetesimal precession rate $A|_{\rm 6AU}$ at $a_p=6$ AU. This choice is almost entirely arbitrary, and is motivated only by the expectation of the reduced disk eccentricity beyond this radius. Precession of the binary is likely to be an insignificant driver of the disk precession outside of $a_A$, where the disk gravity dominates.  For our fiducial system parameters, $A|_{6 {\rm AU}} = -1.2 \times 10^{-3}\, {\rm yr^{-1}}$.

Figure \ref{DiskDominatedErRun} shows that $e_r$ becomes independent of planetesimal size far from the binary, where $e_c$ is low.  This is due to apsidal alignment of planetesimals with the sizes shown in this Figure, as all of them are larger than the critical size $d_c$ in the outer disk, simply because disk eccentricity is very low at large $a_p$, see Equation \eqref{eq:dc}.

%%%%%%%%%%%%%%%%%%%%%%%%%%%%%%%%%%%%%%%%%%%%%

\subsubsection{Secular resonances}
\label{sect:secreswithprec}

Equation \eqref{eq:ec} indicates that disk precession gives rise to two special locations in the disk. First, at the semi-major axis where 
\ba
A=\dot \varpi_d,
\label{eq:disk_res}
\ea 
there is a secular resonance where $e_c \rightarrow \infty$.  This divergence happens because the relative precession between the planetesimal orbits and disk apsidal line vanishes, while the torque exerted by the non-axisymmetric component of the disk gravity is active. As a result, eccentricity can grow without bound in the absence of gas drag. This resonance is an obvious generalization of the secular resonance discussed in \S \ref{sect:fixed_ep}. 

\begin{figure}
\centering
\includegraphics[scale = .5]{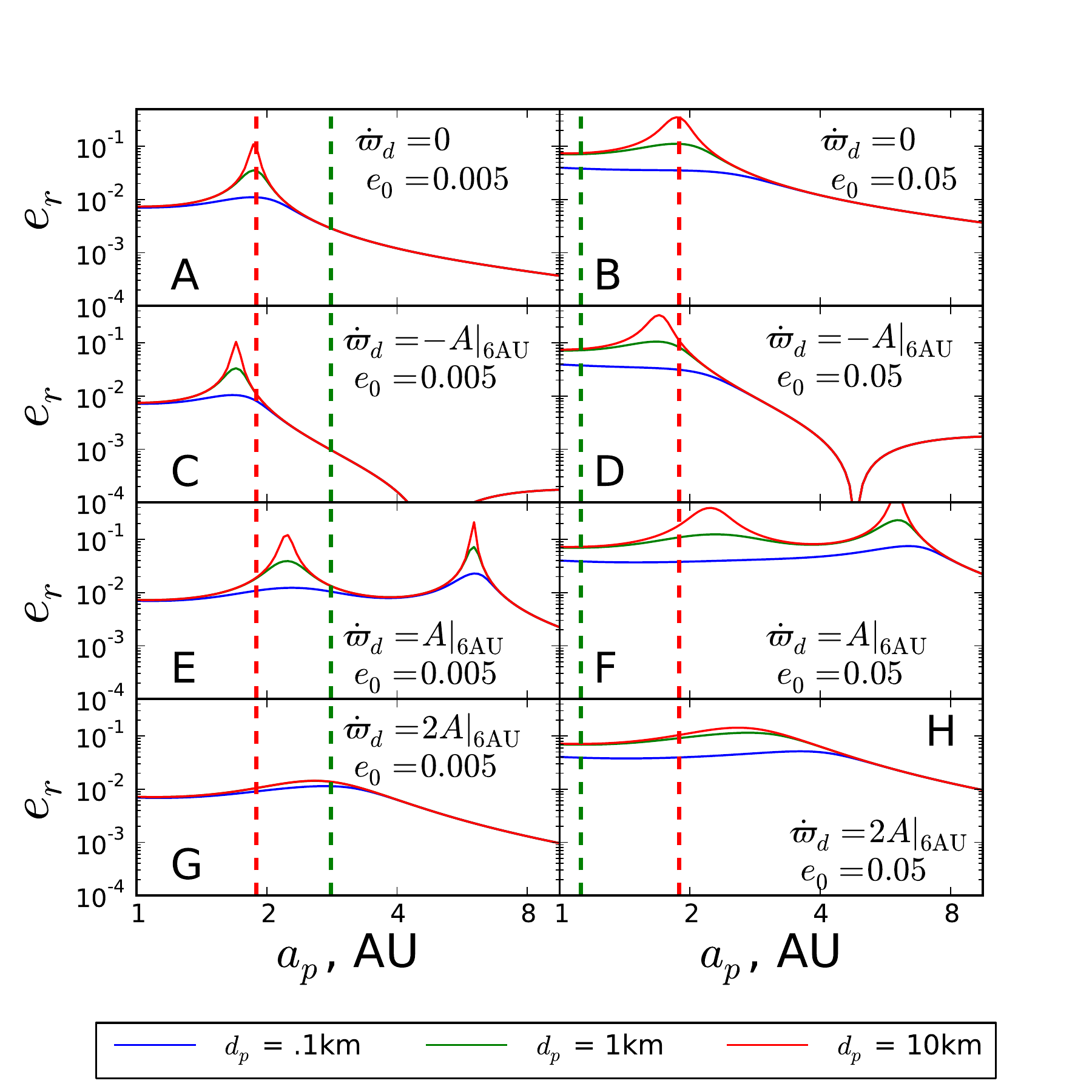}
\caption{Run of relative planetesimal-gas eccentricity $e_r$ in the disk dominated excitation regime, calculated for eight different disk models. We have assumed the fiducial system parameters except where noted. All models on the right have $e_0=0.05$, while all models on the left have lowered disk eccentricity, $e_0=0.005$. Different panels correspond to different disk precession rates in units of $A|_{\rm 6AU}$ (also indicated in Figure \ref{resonances}) --- the planetesimal precession rate at 6 AU: $\dot\varpi_d=0$ (A,B), $\dot\varpi_d=-A|_{\rm 6AU}$ (C,D), $\dot\varpi_d=A|_{\rm 6AU}$ (E,F), $\dot\varpi_d=2A|_{\rm 6AU}$ (G,H). Depending on the disk precession direction and rate curves of $e_r$ feature zero, one, or two secular resonances.}
\label{DiskDominatedErRun}
\vspace{-.05cm}
\end{figure}

Figure \ref{resonances} illustrates the non-trivial behavior of $A$, as it is a combination of $A_d$, and $A_b$, which have opposite signs. We see that depending on the disk precession rate, there can be zero, one, or two secular resonances \eqref{eq:disk_res} associated with disk gravity.  

If $\dot \varpi_d = 0$ (non-precessing disk, yellow level in Figure \ref{resonances}), then there is only one resonance at the location where $A_d + A_b = 0$ (i.e. at 1.9 AU), and the situation is analogous to \S \ref{sect:fixed_ep}, see Figure \ref{DiskDominatedErRun}A,B. As in Figure \ref{erRun}, the larger planetesimals are excited to higher eccentricities at the resonance because they are less damped by the gas drag (see Equation \eqref{er}).

Prograde precession gives rise to similar behavior, as shown in Figure \ref{DiskDominatedErRun}C,D. This is not surprising since Figure \ref{resonances} shows that prograde precession (green level) simply shifts the resonance location inwards, closer to the binary.  This is indeed reflected in Figure \ref{DiskDominatedErRun}C, D, where the resonance is now closer to the star.

As demonstrated in Figure \ref{resonances}, retrograde precession can either remove the resonance if it is very rapid, $|\dot\varpi_d|>|$min$(A)|$ (blue level), or give rise to {\it two} secular resonances as illustrated by the red level. Figure \ref{DiskDominatedErRun}G,H illustrates the former possibility. It is obvious that $e_r$ does not exhibit sharp features in this case. The mild bump around 3 AU is due to the reduced $|A-\dot\varpi_d|$ near the minimum of $A(a_p)$. 

Finally, Figure \ref{DiskDominatedErRun}E,F shows the case of slower retrograde precession, with two conspicuous secular resonances at $a_p\approx 2.2$ AU and  $a_p=6$ AU, in agreement with our expectations. It is clear that in this case $e_r$ can stay at a high level within an extended disk region between the two resonances, harming the prospects for planetesimal growth there.

%%%%%%%%%%%%%%%%%%%%%%%%%%%%%%%%%%%%%%%%%%%%%

\subsubsection{Regions of low $e_r$}
\label{sect:tranquility}

A second type of special location in the disk is possible when the disk precesses in a prograde sense, i.e. in the direction opposite to the precession of the planetesimal orbits. Equation (\ref{eq:ec}) predicts that $e_c \rightarrow 0$ when $\dot \varpi_d = A + B_d/e_d$. When this condition is fulfilled, relative velocities of particles of any size with respect to gas (and, consequently, also w.r.t. each other) vanish at this location, naturally promoting growth.  This cannot occur without disk precession, since $B_d/e_d$ is always larger in magnitude than $A$, except very close to the central binary where $A$ has the same sign as $B_d/e_d$.  

Using equations \eqref{eq:Ad} and \eqref{eq:Bd}, we see that $e_c = 0$ when $\dot \varpi_d = A_b+A_d(1+ \psi_2/2\psi_1)$, and outside $a_A$ the $A_b$ term can be neglected. Since, by our assumption, $\dot \varpi_d$ is constant with radius, but $A_d<0$ is not, one finds that in general a {\it prograde} disk precession ($\dot \varpi_d>0$) will give rise to a radius in the disk where $e_c = 0$, since $\psi_2/\psi_1 \approx -3$. 

This situation is clearly seen in Figure \ref{DiskDominatedErRun}C,D, where disk precession in the direction opposite to $A$ creates a region around $a_p\approx 5$ AU where $e_r \rightarrow 0$, independent of the disk eccentricity. This ``valley of tranquility'', where relative planetesimal velocities are low, may represent a location where planetesimal growth is naturally promoted.

To summarize this analysis, disk precession can be either helpful or harmful to planet formation depending on its direction and magnitude, and on the location in the disk.

%%%%%%%%%%%%%%%%%%%%%%%%%%%%%%%%%%%%%%%%%%%%%
%%%%%%%%%%%%%%%%%%%%%%%%%%%%%%%%%%%%%%%%%%%%%

\section{Planetesimal Dynamics Around a Precessing Binary}
\label{sect:binaryPrecession}

%%%%%%%%%%%%%%%%%%%%%%%%%%%%%%%%%%%%%%%%%%%%%

One important difference between the dynamical environments of P-type and S-type binary systems is that in the P-type systems, the precession rate of the binary itself can easily be comparable to or faster than the planetesimal precession rates for much of the disk \citep{R13}.  It is therefore important to consider the precession of the central binary in calculating planetesimal eccentricities in situations where excitation due to the binary is important.  To that effect, we discuss different mechanisms driving binary precession in Section \ref{sect:generalbinpre} and Appendix \ref{sectbinprecess}. Based on that, we then explore the role of binary precession using a simple disk model in Section \ref{sect:axisymmetricdiskdynamics}.

%%%%%%%%%%%%%%%%%%%%%%%%%%%%%%%%%%%%%%%%%%%%%

\subsection{Binary Precession Rates}
\label{sect:generalbinpre}

%%%%%%%%%%%%%%%%%%%%%%%%%%%%%%%%%%%%%%%%%%%%%

We consider four drivers of binary precession: (1) general relativistic precession, (2) the quadrupole due to tidal interaction between the two stars, (3) the quadrupole due to stellar rotation, and (4) the gravity of the circumbinary disk.  In Appendix \ref{sectbinprecess}, we provide estimates of the magnitude of each contribution.

We find that precession due to the tidally induced quadrupole is dominant for massive stars that are close together, and disk-driven precession is dominant for less massive and more widely separated binaries.  General relativistic precession and rotationally induced quadrupole precession are always subdominant to one or the other of tidal or disk precession for the Kepler systems. 

\begin{figure}
\centering
\includegraphics[scale = .38]{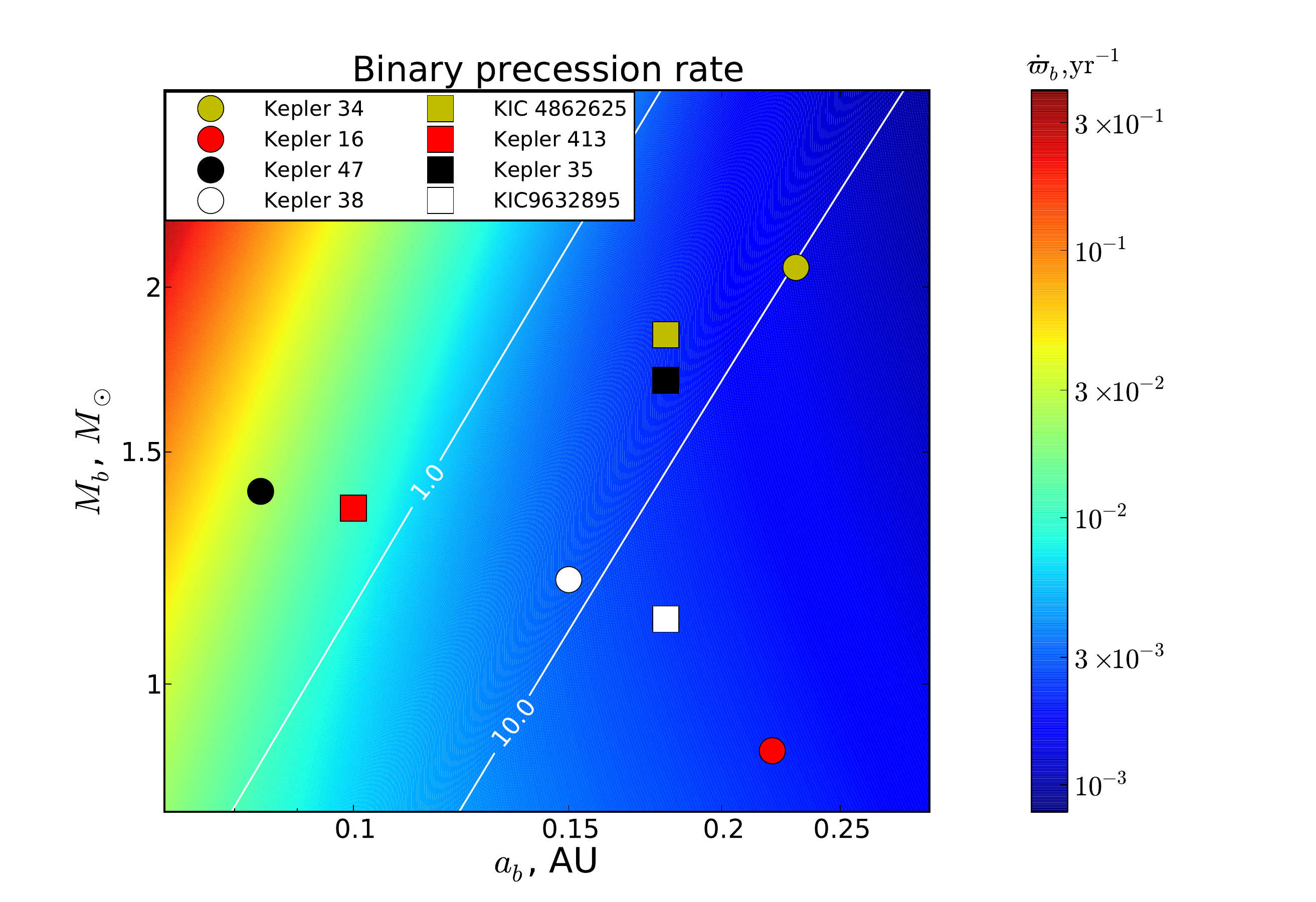}
\caption{Binary precession rate (yr$^{-1}$), as a function of $a_b$ and $M_b$. Thin white lines show the contours where disk precession equals tidal precession, and where it exceeds the latter by a factor of 10.  Points correspond to known circumbinary systems listed in Table \ref{table:sysParams}.  We have assumed $\mu = 1/3$ in calculating the precession rate. See text for more details.}
\label{binaryPrecessionRate}
\vspace{-.05cm}
\end{figure}

Figure \ref{binaryPrecessionRate} shows our estimate of the binary precession rate as a function of $M_b$ and $a_b$.  This assumes that $\mu = 1/3$, the apsidal motion constant $k_2 = 0.13$ (see Equations \eqref{tpr} and \eqref{rpr}) and the log of the surface gravity (shown in the work of \citet{Cl12} to be nearly constant over a wide mass range for 1 Myr old stars) is 3.66 in cgs units.  We have taken these numbers from the pre-main-sequence stellar models of \citet{Cl12} for stars 1 Myr in age. We assume our fiducial disk model (Table \ref{table:fidParams}) in calculating the contribution of disk gravity to binary precession.  

We have additionally placed on this figure eight Kepler binary systems known to host circumbinary planets.  The precession rates for these systems are only approximate, as they do not all have $\mu = 1/3$ assumed in our calculation, and there is a substantial change in stellar radii over the course of pre-main-sequence evolution.  The contours correspond to locations in $a_b-M_b$ space where disk precession rate equals tidal precession rate, and where it exceeds the latter by a factor of 10.  If the disk cavity were larger (i.e. $a_{\rm in}$ were larger than $2a_b$), then the disk would drive substantially slower binary precession, see Appendix \ref{sectbinprecess}.

We see from Figure \ref{binaryPrecessionRate}, that assuming precession to be disk-dominated is a good approximation for most of these systems, and particularly for Kepler 16, for which binary precession is dominated by the disk by a factor of hundreds over the tidal quadrupole precession.  Throughout the rest of the paper, we assume that binary precession is solely due to the disk.

Using Equations \eqref{eq:Ad} and \eqref{binaryPrecessionFromDisk}, with $\psi_1 = -0.55$, we find
\begin{equation}
\label{eq:Advarpid}
\frac{\dot \varpi_b}{A_d} =  -0.15 \frac{a_p}{a_{\rm in}}
\end{equation}
for our disk model with $p = 1.5$.  Equation \eqref{binaryPrecessionFromDisk} yields $\dot \varpi_b \sim a_{\rm in }^{-1}$ because we are assuming $a_b/a_{\rm in} = 1/2$, and $A_d \sim a_p^{-1}$ from Equation \eqref{eq:Ad}.  Because $A_d$ and $A_b$ have different signs, $|A|<|A_d|$. As a result, even inside of the radius where $A_d = \dot \varpi_b$, the major contribution to the relative precession between the binary and the planetesimals may be coming from the binary precession --- see Figure \ref{resonances}.

%%%%%%%%%%%%%%%%%%%%%%%%%%%%%%%%%%%%%%%%%%%%%

\subsection{Axisymmetric Disk}
\label{sect:axisymmetricdiskdynamics}

%%%%%%%%%%%%%%%%%%%%%%%%%%%%%%%%%%%%%%%%%%%%%

As an application, here we consider planetesimal dynamics in the limiting case of an {\it axisymmetric} disk ($e_0=0$) with a central binary precessing at the rate $\dot \varpi_b$.  This section considers the same setup as in \citet{R13}, but with the addition of gas drag. 

Setting $\varpi_b = \dot \varpi_b t$ and $e_d\to 0$, $B_d\to 0$ in Equations (\ref{DDEquations1})-(\ref{DDEquations2}) we can easily solve them analytically. We find that the solution for ${\bf e}_p$ is identical to Equation (\ref{alignedSolution}) if we set $e_d\to 0$ and replace $B_d\to B_b$, $\dot \varpi_d\to \dot \varpi_b$ in the latter.  In this solution, the forced eccentricity vector ${\bf e}_p$ rotates at the rate $\dot\varpi_b$ and is fixed in the frame precessing with the binary. 

The relative planetesimal-gas eccentricity is still given by Equation \eqref{er} but with $\dot \varpi_d$ replaced by $\dot \varpi_b$. And instead of Equation \eqref{eq:ec} we now have
\begin{equation}
\label{axisymec}
 e_c = |\frac{B_b}{A -\dot \varpi_b}|.
\end{equation}

Given that binary precession is prograde, $\dot \varpi_b>0$, Figure \ref{resonances} implies that there could be only one secular resonance associated with binary precession (illustrated by the orange level calculated for our fiducial parameters of Kepler-16) and determined by the condition 
\ba
A=\dot \varpi_b.
\label{eq:bin_res}
\ea 
It is located somewhat closer to the star than $a_A$. Given all that, it is clear that the plot of $e_r$ is going to be similar to Figure \ref{DiskDominatedErRun}C,D. However the ``valley of tranquility'' is going to be absent, and the secular resonance would now correspond to the location where the condition \eqref{eq:bin_res} is satisfied.
\par
In the case of very tight binaries ($a_b \lesssim 0.1$AU) binary precession is much faster than the planetesimal precession rate $A$ throughout the disk, thus strongly suppressing the excitation due to the binary. This effect was first noticed in \citep{R13}.  Even more modest precession rates, such as those calculated for Kepler-16 in this paper ($\dot \varpi_b = 2.6 \times 10^{-3} \, {\rm yr^{-1}}$) reduce the excitation effect of the binary by factors of a few.  Looking at Figure \ref{resonances}, we see that in most of the disk outside $a_A$, the main contribution to the relative planetesimal-binary precession $A - \dot \varpi_b$ is due to the high value of $\dot \varpi_b$. Thus, binary precession effectively suppresses planetesimal eccentricity excitation due to its own non-Newtonian potential in the outer parts of the disk.

%%%%%%%%%%%%%%%%%%%%%%%%%%%%%%%%%%%%%%%%%%%%%
%%%%%%%%%%%%%%%%%%%%%%%%%%%%%%%%%%%%%%%%%%%%%

\section{Dynamics With Both Binary and Disk Gravity and Both Binary and Disk Precession}
\label{sect:gen_dyn}

%%%%%%%%%%%%%%%%%%%%%%%%%%%%%%%%%%%%%%%%%%%%%

To conclude our discussion of circumbinary planetesimal dynamics we provide a general description of the ${\bf e}_p$ behavior in the most general situation when neither disk or binary gravity, nor disk or binary precession can be ignored, and {\it quadratic} gas drag damps planetesimal eccentricity. In this case the forced planetesimal eccentricity is no longer constant in time (even in some precessing frame) and its amplitude varies. As shown in \citet{Beauge10} and SR15, evolution of ${\bf e}_p$ can be viewed as a superposition of the two precessions in the eccentricity space, resulting in a limit cycle behavior with $d_p$-dependent characteristics, which cannot be described analytically. 

Despite this complication, important insights into the general problem can be gained by analyzing the general solution for the {\it linear} gas drag law described in Appendix \ref{sect:lindrag}. Examination of this solution shows that in general one should expect secular resonances of both types --- given by Equations (\ref{eq:disk_res}) and (\ref{eq:bin_res}) --- to exist in the disk. Given the discussion in \S \ref{sect:secreswithprec} and \ref{sect:axisymmetricdiskdynamics} one expects up to three secular resonances to emerge. In particular, three resonances appear for slow retrograde disk precession, with two resonances being due to disk gravity (corresponding to the situation in Figure \ref{DiskDominatedErRun}E,F) and one due to the binary precession (see Equation (\ref{eq:bin_res})). This makes planetesimal dynamics even more complex than before. 

Nevertheless, the general features of the $e_r$ behavior outlined in \S \ref{sect:fixed_ep} and \ref{sect:diskPrecess} and shown Figures \ref{erRun} and \ref{DiskDominatedErRun} remain in place: eccentricity reaches high values at resonances, with the larger increase of $e_r$ for bigger objects, less coupled to gas. At large separations disk gravity would still dominate and drive ${\bf e}_p$ to size-independent behavior, see \S \ref{sect:np} and \ref{sect:secreswithprec}. Lower disk eccentricity $e_0$ would still result in lower planetesimal eccentricity, and so on. 

This completes our discussion of secular planetesimal dynamics in circumbinary disks.

%%%%%%%%%%%%%%%%%%%%%%%%%%%%%%%%%%%%%%%%%%%%%
%%%%%%%%%%%%%%%%%%%%%%%%%%%%%%%%%%%%%%%%%%%%%

\section{Collisional Outcomes and Growth of Planetesimals}
\label{sect:CollisionalOutcomes}

%%%%%%%%%%%%%%%%%%%%%%%%%%%%%%%%%%%%%%%%%%%%%

To assess the prospects for planet formation in circumbinary systems, we must couple our understanding of the planetesimal dynamics outlined in previous sections with a prescription for the outcome of planetesimal collisions. We adopt that from RS15b, who based their calculation on the results of \citet{SL09}. In their framework the outcome depends on the masses of the planetesimals, an assumption about their internal strength, and their collision velocity, which is determined by the ${\bf e}_p$ of each body involved in a collision.  

For completeness, details of our collisional prescription are reproduced in Appendix \ref{sect: collOut}.  We categorize collisions in three groups based on the mass of the largest surviving fragment.
 
\begin{itemize}
 
\item {\it Catastrophic} collisions leave a largest remnant containing no more than half the combined mass of the two colliding bodies. 
  
\item An {\it erosive} collision leaves a largest remnant smaller than the larger of the two incoming planetesimals.  

 \item If the largest remnant is bigger than either of the two incoming bodies, we say that the collision leads to {\it growth}. 
 
 \end{itemize}
 \par

\begin{figure}
\centering
\includegraphics[scale = .3]{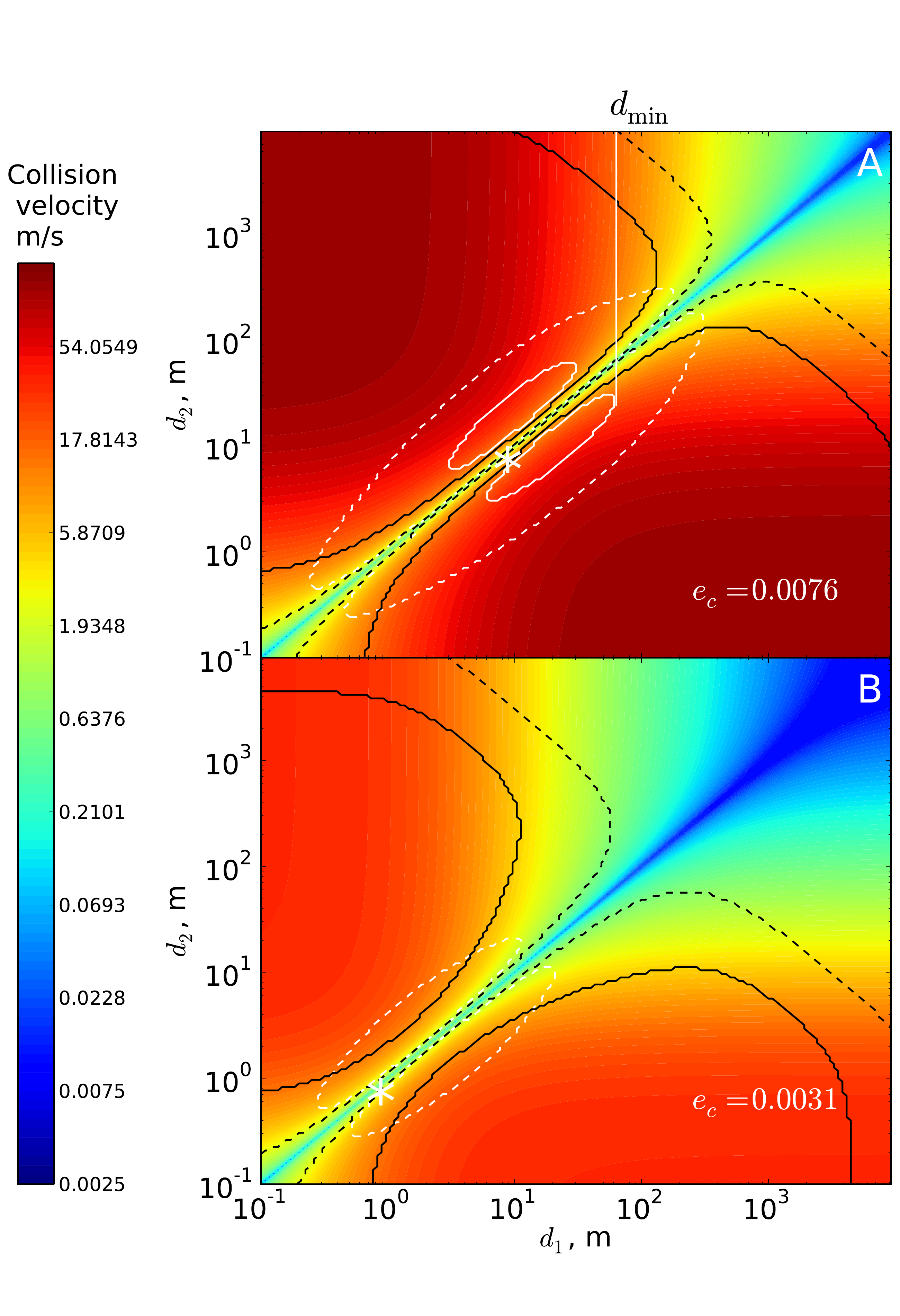}
\caption{Regions of catastrophic disruption and erosion for two different environments in the space of sizes of the two colliding partners.  Panel A is made for our fiducial system at 3.5 AU with no disk or binary precession, assuming dynamics derived in \S \ref{sect:diskdominatedregimedynamics}.  White lines enclose regions of destructive collisions for ``strong" (solid) and ``weak" (dashed) planetesimals \citep{SL09}.   Black lines enclose regions of erosive collisions for strong (solid) and weak (dashed) planetesimals.  Panel B shows the same system at 6 AU.  Here we see that although there is {\it no} destruction region for strong planetesimals, a large range of collisions still result in erosion.  We have also shown the value of $e_c$ and placed a white star at the locations of $d_c$ in each panel. Definition of the critical planetesimal size necessary for growth $d_{\rm min}$ is illustrated in the top panel.}
\label{erosionFig}
\vspace{-.05cm}
\end{figure}

Figure \ref{erosionFig} illustrates collisional outcomes in the space of sizes of colliding planetesimals $d_1$ and $d_2$ at two different locations in the circumbinary disk of the Kepler 16 system, using the disk-dominated excitation approximation without precession, and with fiducial disk parameters.  White contours enclose regions leading to catastrophic disruption.  Black contours enclose regions leading to erosion.  

We see that collisions of bodies of exactly the same size lead to growth because collision velocities are small.  Catastrophic disruption occurs only closer to the binary (panel A), since $e_c$ is not high enough further out (panel B) for catastrophic collisions to occur.  We see that similar in size, but not exactly equal planetesimals tend to undergo catastrophic collisions, because their ${\bf e}_p$ are different enough to result in significantly energetic collisions.  However, even very unequal mass ratio collisions can lead to erosion. 
\par
The white star in each plot corresponds to the size $d_c$.  We notice that this is near the region of catastrophic destruction in the plot.  The lobes are biased towards $d_p > d_c$ because $d_c$ is smaller than the size of planetesimal with the lowest critical velocity for destruction, which is about 100 m, see Figure 2 of RS15b.  Because the destruction region is near $d_c$, having a low value of $d_c$ means that planets can form starting from a population of smaller planetesimals. In fact, Figure \ref{erosionFig} suggests that if the planetesimal population were to contain only the objects with sizes $\gtrsim (30-50)d_c$, this population would have been completely immune to both catastrophic disruption and erosion.

In this paper, we consider catastrophic disruption to be the only obstacle to planetesimal growth.  For a number of reasons discussed further in Section \ref{sect:erosion}, we do not expect erosion to play a determining role in whether planetesimal growth occurs. 

In the following sections we examine the prospects for collisional growth in each of the dynamical regimes discussed in Sections \ref{sect:exactNoDiskPrecess} - \ref{sect:binaryPrecession}, which provide different prescriptions for $e_c$ and $d_c$.  Using these, we calculate the {\it smallest planetesimal size} $d_{\rm min}$ such that objects larger than $d_{\rm min}$ do not suffer catastrophic disruption.  This size is indicated in panel A of Figure \ref{erosionFig}.  This is the smallest planetesimal size that we can start from and grow larger objects without ever encountering catastrophic disruption. Because of our neglect of erosion, overall planetesimal growth requires only the existence of objects larger than $d_{\rm min}$ in the planetesimal populations.

If this size is under 10 m, we conclude that planetesimal growth via collisional agglomeration is easy under those environmental conditions.  The choice of 10 m is somewhat arbitrary, but we note that changing from 10 to 100 m in our plots would make very little difference to our conclusions.  If $d_{\rm min} > 10$ m, $d_{\rm min}$ provides an estimate of the minimum size of primordial planetesimals necessary to form planets in that environment.  Unless otherwise specified, we are assuming the material properties appropriate for the ``strong planetesimals" of \citet{SL09}.

%%%%%%%%%%%%%%%%%%%%%%%%%%%%%%%%%%%%%%%%%%%%%
%%%%%%%%%%%%%%%%%%%%%%%%%%%%%%%%%%%%%%%%%%%%%

\subsection{Planetesimal growth in the Disk-Dominated Excitation Regime} 
\label{sect:pgdder}

%%%%%%%%%%%%%%%%%%%%%%%%%%%%%%%%%%%%%%%%%%%%%

\begin{figure}
\centering
\includegraphics[scale = .5]{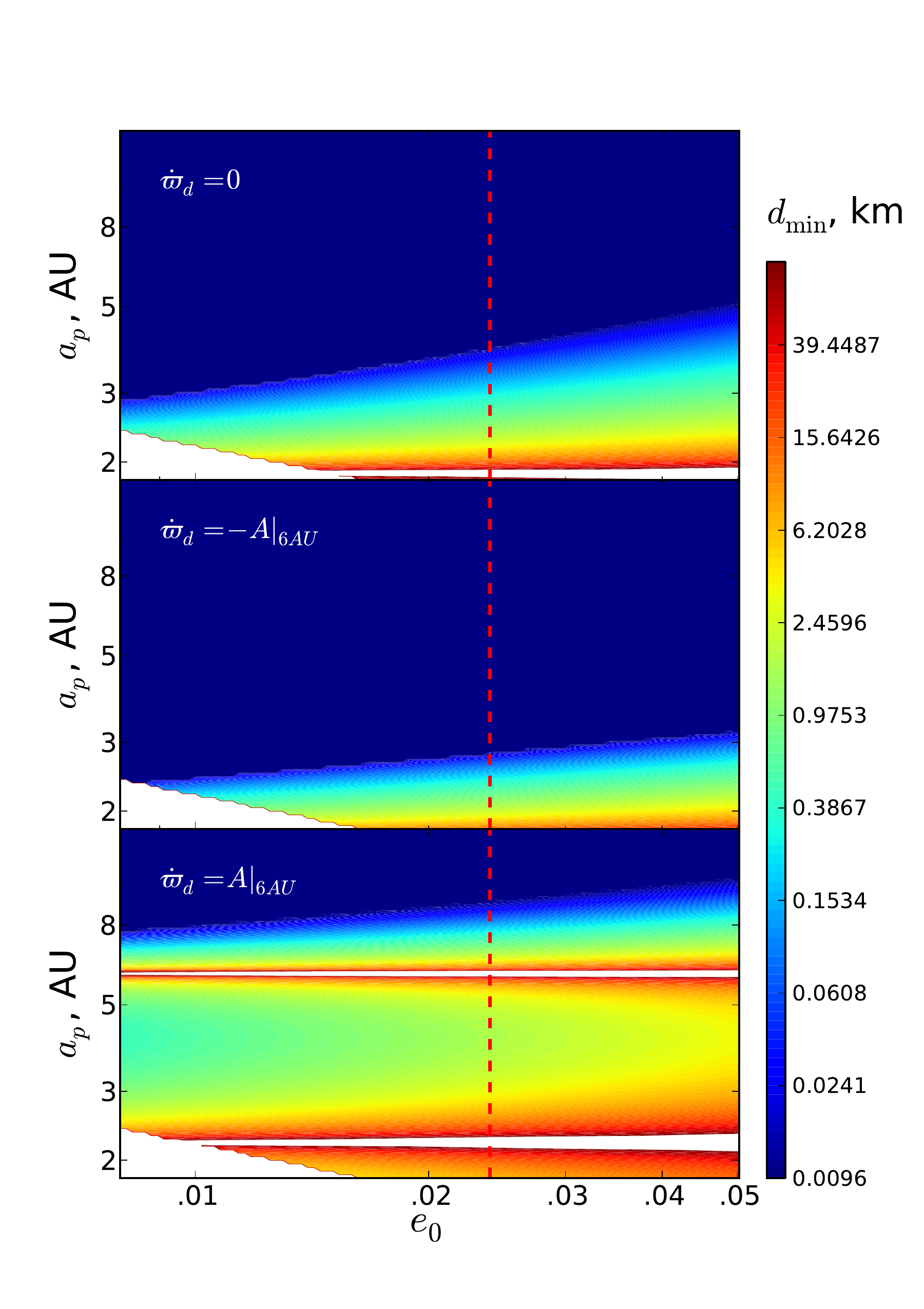}
\caption{Minimum planetesimal size $d_{\rm min}$ safe from catastrophic disruption in the disk dominated excitation regime (described in \S \ref{sect:diskdominatedregimedynamics}).  Calculation is done for our fiducial disk around Kepler 16, see Table \ref{table:fidParams}. Different panels correspond to different assumptions about the disk precession rate and direction.  The whited out region in the bottom left is where $|B_b| > |B_d|$ and we are not justified in using the disk dominated approximation.  The dashed red line is drawn at the value of $e_0$ corresponding to the forced eccentricity of a free particle in the binary potential, see Equation (\ref{eq:eforced}).}
\label{diskPlusBinary}
\vspace{-.05cm}
\end{figure}

We first consider the collision outcomes in the outer part of the disk where excitation from the binary is unimportant compared with excitation from the disk, and use the results on planetesimal dynamics from Section \ref{sect:diskdominatedregimedynamics}.  

Figure \ref{diskPlusBinary} shows the size $d_{\rm min}$, above which the growth is unimpeded by catastrophic disruption, as a function of semi-major axis and disk eccentricity for different assumptions about the disk precession.  We have used our fiducial system parameters listed in Table \ref{table:fidParams}.  Collision outcomes have been calculated using Equations \eqref{eq:ec} and \eqref{eq:dc} for $e_c$ and $d_c$. 

One can see that in general, the outer region of the disk is favorable for growth, even starting from small ($d_p \sim$ 10\, m) planetesimals.  This is true for two reasons.  First, local disk eccentricity is small at large $a_p$, which leads to low values of $e_c$.  Second, far from the binary $d_c \propto a_p^{-13/4}$, and becomes quite low (around a meter) at around 5 AU.  This means that all the planetesimals have $(A - \dot \varpi_d) \tau_d \gg 1$ and $e_r$ very near $e_c$.

%%%%%%%%%%%%%%%%%%%%%%%%%%%%%%%%%%%%%%%%%%%%%

\subsubsection{Effects of Disk Precession}

Looking at Figure \ref{diskPlusBinary}, we see that in the absence of disk precession, the outer region of the disk is friendly to planetesimal growth beyond 3-4 AU (for all planetesimal sizes down to 10 m) depending on $e_0$.  Starting with planetesimal sizes of a few km brings the inner edge of the growth-friendly region within 3 AU, even for disks with $e_0 = 0.05$ (twice the free-particle eccentricity).  

In the middle panel the growth region expands because prograde disk precession ($-A|_{\rm 6AU}$ is positive) dramatically lowers $e_c$.  In this case, the disk precesses at $\dot \varpi_d = A|_{6 AU}$ ($A|_{\rm 6 AU}$ is the planetesimal precession rate at 6 AU), and we find nearly the whole outer disk ($a_p>3$AU) to be conducive to planetesimal coagulation even for highly eccentric disks.  This is because of the valley of low $e_c$ in a prograde precessing disk, discussed in Section \ref{sect:diskPrecess} and shown in Figure \ref{DiskDominatedErRun}C,D.

On the other hand, retrograde disk precession may give rise to a second secular resonance, which is very damaging to planetesimal growth, see Figure \ref{diskPlusBinary}C.  This resonance makes conditions in the outer part of the disk, around 6 AU, much more hostile to planetesimal growth than in the absence of disk precession. In addition, comparing panels A and C, we see that the inner resonance in panel C is also moved slightly outwards because of the disk precession. As a result, planetesimals within a broad radial interval of the disk (1-8 AU) end up being strongly dynamically excited.

%%%%%%%%%%%%%%%%%%%%%%%%%%%%%%%%%%%%%%%%%%%%%

\subsubsection{Effect of Planetesimal Strength}

\begin{figure}
\centering
\includegraphics[scale = .5]{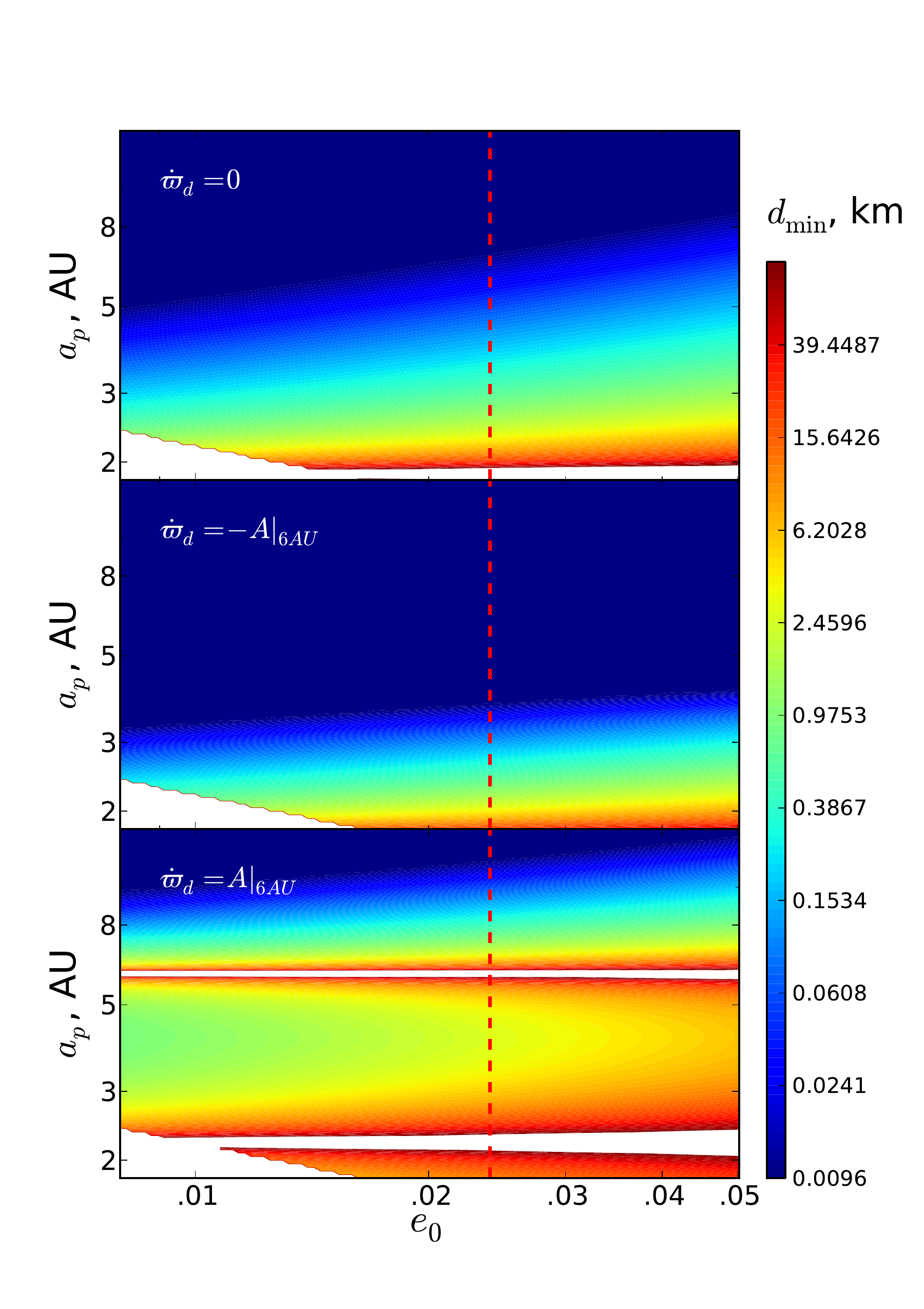}
\caption{Same as Figure \ref{diskPlusBinary} except that we are using the collisional prescription for weak aggregates \citep{SL09} instead of that for solid rocks, see Appendix \ref{sect: collOut}.}
\label{diskPlusBinaryWeak}
\vspace{-.05cm}
\end{figure}

Next, we consider changing the material properties of the planetesimals, i.e. changing the $Q^*_{RD}$ term in Equation (\ref{eq:SL09}).  \citet{SL09} provide two prescriptions for $Q^*_{RD}$ depending on whether planetesimals are assumed to be solid rocks (strong planetesimals), or rubble piles (weak planetesimals).  Figure \ref{diskPlusBinaryWeak} shows the same dynamical environment as Figure \ref{diskPlusBinary}, but considers weak planetesimals.  This does not make a big difference except for the sub-kilometer sized planetesimals, as there is not a big difference between weak and strong planetesimals in the gravity-dominated regime (for $d_p\gtrsim 1$ km).  Comparing Figures,  \ref{diskPlusBinary} and \ref{diskPlusBinaryWeak}, we see differences of several AU in the extent of the region where 10 m sized planetesimals are vulnerable to catastrophic disruption, but insignificant differences in the extent of the region where km-sized planetesimals are subject to destruction.

%%%%%%%%%%%%%%%%%%%%%%%%%%%%%%%%%%%%%%%%%%%%%

\subsubsection{Disk Mass}
\label{sect:disk mass outcomes}

\begin{figure}
\centering
\includegraphics[scale = .5]{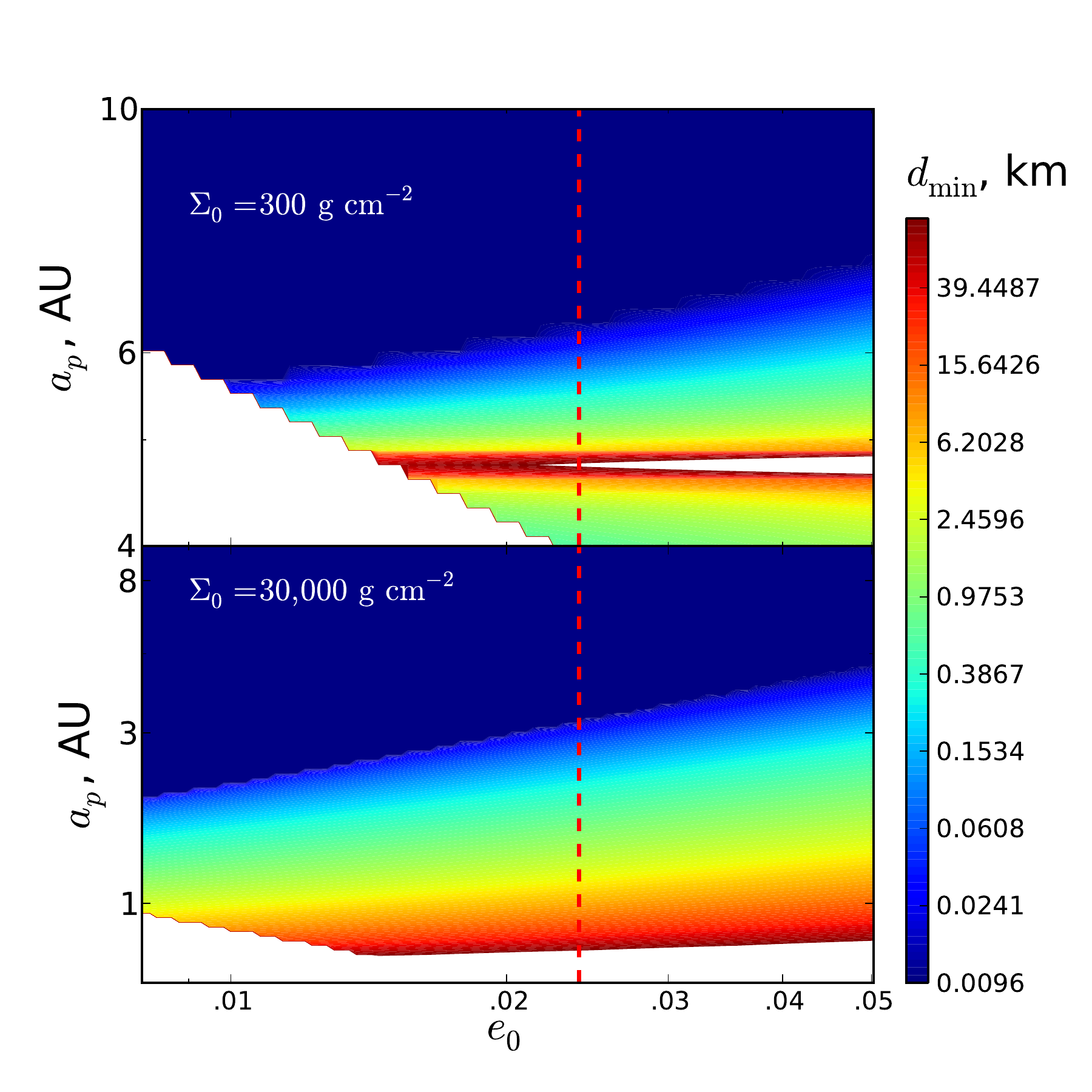}
\caption{Same as the top panel of Figure \ref{diskPlusBinary} but now for two different disk masses, resulting in different surface densities at 1 AU $\Sigma_0$, indicated on panels. We use the system parameters in Table \ref{table:fidParams} except for the disk density. }
\label{diskPlusBinaryLight}
\vspace{-.05cm}
\end{figure}

Another model parameter which we vary is the surface density at 1 AU.  Unlike the case of tight S-type systems, there is little reason to believe that circumbinary disks contain less mass than their counterparts around single stars.  In fact, if anything, there seems to be evidence for more massive disks in circumbinary systems \citep{Harris12}.  

In Figure \ref{diskPlusBinaryLight} we consider both a denser and less dense disk than our fiducial system, with $\dot \varpi_d = 0$, again using the dynamics discussed in Section \ref{sect:diskdominatedregimedynamics}.  We have adjusted the scale on the $y$-axis of Figure \ref{diskPlusBinaryLight} in both subplots so that the bottom of the plot is near $a_B$, where excitation switches to being dominated by the disk, see Equation \eqref{eq:bequality}.  Because $a_B$ depends on disk mass, the scales on the $y$-axes are not the same in the two panels.  We have whited out the area in the plot where $|B_b| > |B_d|$, as we do not have an analytic solution for $e_c$ in that case.  

The main effect of changing the disk mass is the variation in location of the secular resonance where $A_d + A_b=0$: it moves out for lower $\Sigma_0$ (and $M_d$). For this reason lowering the disk mass is quite unfavorable for planet formation in the outer part of the disk. There may however be a region favorable to growth interior to the resonance, a possibility we explore further in Section \ref{sect:fastStop}.

For the very massive disk with $\Sigma_0 = 3 \times 10^4$ g cm$^{-2}$, we find that the dynamics look similar to the top panel of Figure \ref{diskPlusBinary} (again accounting for the change in scale on the $y$-axis). This is because in the disk dominated excitation regime, $d_c$ and $e_c$ are independent of disk mass, see discussion after Equation (\ref{eq:dc}).  The only difference we expect from increasing the disk mass is that the secular resonance moves inward.  This is why the differences between Figures \ref{diskPlusBinary} and \ref{diskPlusBinaryLight} become more striking as one moves inwards in semi-major axis.  

%%%%%%%%%%%%%%%%%%%%%%%%%%%%%%%%%%%%%%%%%%%%%

\subsection{Collisional Outcomes in an Axisymmetric Disk with Binary Precession}
\label{sect:axisymmOutcomes}

%%%%%%%%%%%%%%%%%%%%%%%%%%%%%%%%%%%%%%%%%%%%%

 As found in \citet{R13}, a massive axisymmetric disk is very helpful for reducing planetesimal collision velocities, both because of the enhanced planetesimal precession and because of the induced binary precession. Not surprisingly, when including gas drag, this result remains valid, as we show now.  

\begin{figure}
\centering
\includegraphics[scale = .5]{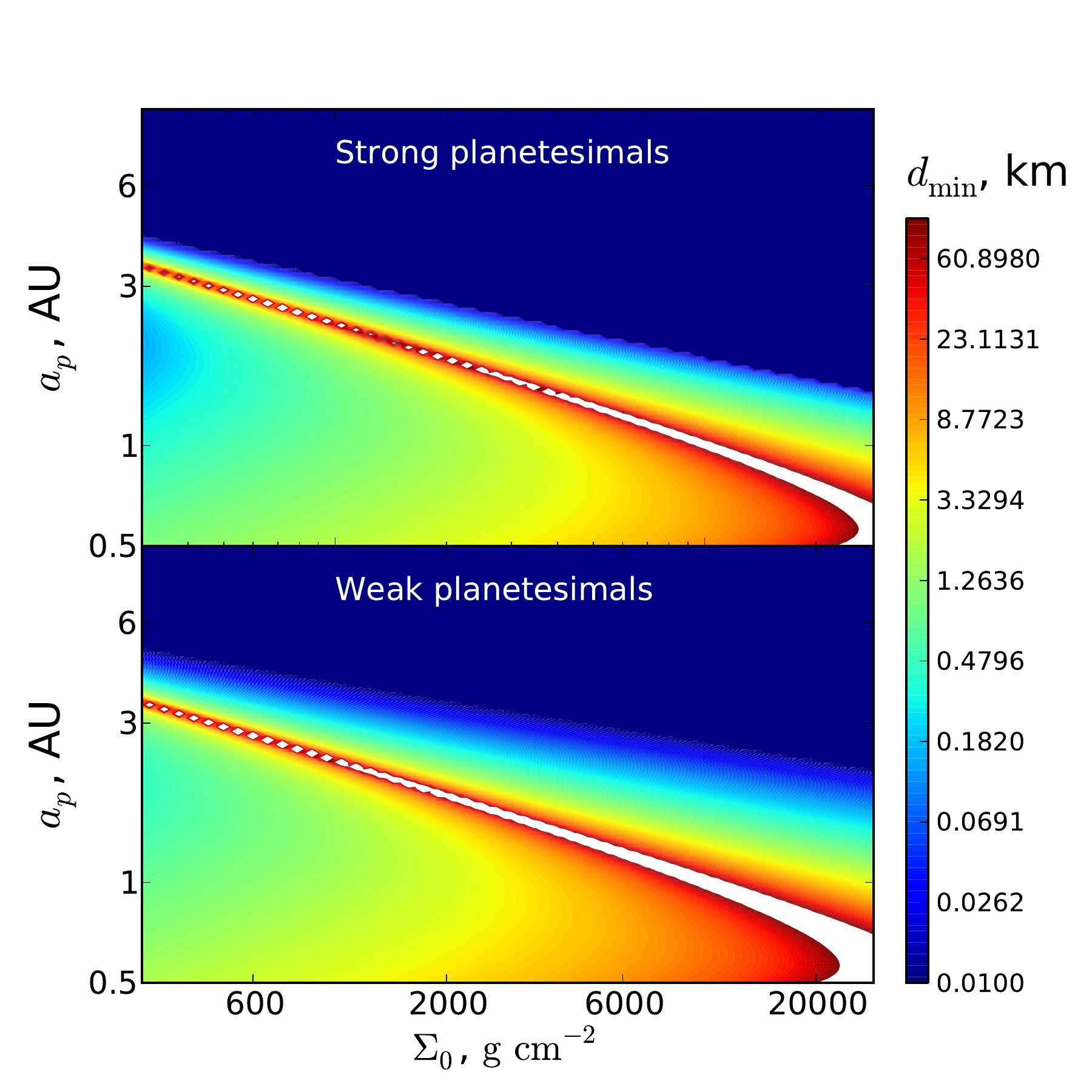}
\caption{Size $d_{\rm min}$, as a function of $\Sigma_0$ and $a_p$ for an axisymmetric disk with $p = 1.5$ in the Kepler 16 system. Planetesimal collisional velocities are calculated using the results of \S \ref{sect:binaryPrecession}. The two panels correspond to the strong and weak planetesimals discussed in \citet{SL09}. Binary precession rate $\dot \varpi_b$ is given as a function of disk mass ($\Sigma_0$) by Equation \eqref{binaryPrecessionFromDisk}.}
\label{axisymmetricSizePlot}
\vspace{-.05cm}
\end{figure}

In Figure \ref{axisymmetricSizePlot} we show $d_{\rm min}$ in the axisymmetric disk approximation as a function of $\Sigma_0$ and $a_p$. We are now using the dynamics discussed in Section \ref{sect:binaryPrecession}, including the effects of non-zero binary precession, in particular Equations \eqref{eq:Advarpid} and \eqref{axisymec}. We consider both strong (top) and weak (bottom) planetesimals.

Both panels exhibit similar structure, since the planetesimal dynamics are the same, independent of their material properties. We see the secular resonance given by Equation (\ref{eq:bin_res}) running diagonally across both panels. Its appearance is different from Figures \ref{diskPlusBinary}, \ref{diskPlusBinaryWeak} and \ref{diskPlusBinaryLight}, because those figures held disk mass constant.  

Exterior to the resonance, where $A$ is disk-dominated, $d_c$ is independent of disk mass, see Equation \eqref{eq:dc}.  However a more massive axisymmetric disk lowers $e_c$ by increasing the rate of relative precession between the binary and planetesimal orbits, without adding to the excitation.  We see that a high $\Sigma_0$ allows unimpeded coagulation as close as 2.5 AU even for weak planetesimals.  A low disk density is detrimental to planetesimal growth at large radii because the secular resonance moves out.   

Interior to the secular resonance, we see that a massive disk is actually harmful to the survival of smaller planetesimals. Indeed, fixing e.g. $a_p=1$ AU and increasing $\Sigma_0$ leads to higher $d_{\rm min}$. This is because in the inner disk $A$ is dominated by the binary, $A \approx A_b$, so that a more massive disk increases both the value of $d_c$ (because $\Sigma_d$ increases in Equation \eqref{eq:dc}), and $e_c$ (by moving the resonance inwards).  

In the opposite limit of a low mass disk ($\Sigma_0$ = 300 g cm$^{-2}$ corresponding to a disk mass of a few Jupiter masses), in situ planetesimal growth at $a_p \approx 1 {\rm AU}$ is possible with km-sized initial planetesimals.  A light disk ensures that the secular resonance is at several AU, and therefore not playing a role in the dynamics inside of an AU.  It also means that $d_c$ is still substantially smaller than a kilometer, even at an AU separation.  For example, for an axisymmetric disk in our fiducial system with $\Sigma_0 = 300$ g cm$^{-2}$, the critical size is just 8 m at 1 AU.

%%%%%%%%%%%%%%%%%%%%%%%%%%%%%%%%%%%%%%%%%%%%%

\subsection{Planetesimal Growth Near the Star in the Limit of Short Stopping Times}
\label{sect:fastStop}

%%%%%%%%%%%%%%%%%%%%%%%%%%%%%%%%%%%%%%%%%%%%%

\begin{figure}
\centering
\includegraphics[scale = .5]{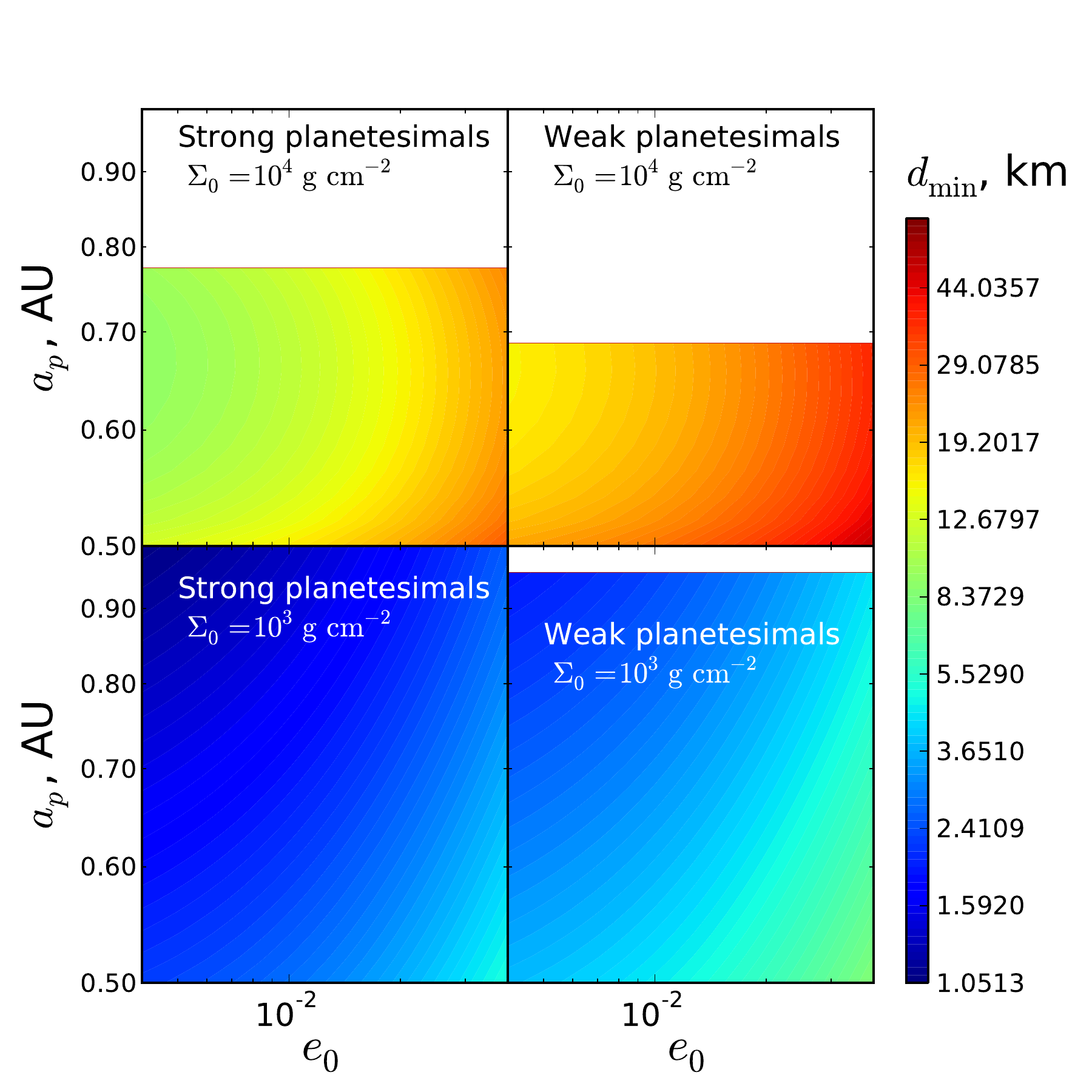}
\caption{Size $d_{\rm min}$ in the inner part of the disk, adopting approximation of no disk and binary precession (\S \ref{sect:exactNoDiskPrecess}), which is valid for small bodies. We calculate $e_c$ by maximizing Equation \eqref{eq:ecRS15a} over $\varpi_d-\varpi_b$.  White areas correspond to locations where Equation \eqref{eq:ecRS15a} is no longer valid because $\dot \varpi_b \tau_d > 1$, for the largest size of planetesimal which is destroyed. System parameters are taken from Table \ref{table:fidParams} except for the disk density.}
\label{innerDisk}
\vspace{-.05cm}
\end{figure}

Planetesimal dynamics are complicated near the star because both gas drag and binary gravity are important, and the disk and binary do not precess at the same rate, causing planetesimal eccentricities to be time dependent, see \S \ref{sect:gen_dyn}. Nevertheless, we still find analytic solutions for planetesimal orbits in two limits. One is the axisymmetric disk discussed in the previous section.  The other is the limit of no disk and binary precession discussed in Section \ref{sect:exactNoDiskPrecess}.  

Planetesimals small enough to have stopping times short compared with the binary and disk precession times (max$(\dot \varpi_b,\dot \varpi_d) \tau_d \ll 1$) should have their eccentricities approximately described by Equations (\ref{eq:vect_sum})-(\ref{eq:phib_nonprec}) with $\varpi_b$, $\varpi_d$ given by the instantaneous orientation of the binary and the disk. We imagine $|\dot \varpi_d|$ to be smaller than $|\dot \varpi_b|$, so we are only requiring $\dot \varpi_b\tau_d \ll 1$, when assessing the validity of the approximation that stopping times are rapid. 

Planetesimal eccentricity (\ref{eq:vect_sum}) obtained in Section \ref{sect:exactNoDiskPrecess} is a function of the mutual disk-binary apsidal orientation $\varpi_d - \varpi_b$. To be conservative in our estimate of the planetesimal destruction region, here we calculate the {\it maximum} value of $e_c$ as a function of $\varpi_d - \varpi_b$ in Equation \eqref{eq:ecRS15a}, thus considering planetesimal dynamics in the least favorable part of the binary orbit.  

We use these assumptions to generate Figure \ref{innerDisk}, which is similar to Figure \ref{diskPlusBinary}. We display as white the region where $\dot \varpi_b \tau_d < 1$ for planetesimals of size $d_{\rm min}$.  This is where we expect the approximation of a slowly-precessing binary to break down.

We see a substantial difference of the outcomes depending on the density of the disk. Because the gravitational perturbations are dominated by the binary (i.e. $A$ is independent of $\Sigma_d$), the critical size $d_c$ is smaller in lighter disks, see Equation (\ref{eq:dc}), leading to even km-sized bodies having more aligned orbits.  Additionally, the secular resonance moves outwards for lower $\Sigma_0$, leading to lower $e_c$ within an AU. 

As a result, in the case of a low-density, low-eccentricity disk, we see that it is possible to have planetesimals greater than a few km in size grow undisturbed by catastrophic disruption.  This is a simpler scenario for planetesimal growth than the one described in \citet{Meschiari2014} which relies on a pressure maximum to trap small dust to enable planetesimal growth.  One caveat of the in-situ growth scenario with a low-$\Sigma_0$ disk is that it may be difficult to grow Saturn-size circumbinary planets (such as Kepler 16b) because of the short supply of mass in such a disk.

For the denser disks (higher $\Sigma_0$), catastrophic disruption is inevitable unless nature creates initial planetesimals on the order of tens of km in size \citep{Johansen12}, see \S \ref{LIPs}.  These general conclusions are similar to those obtained in Section \ref{sect:axisymmOutcomes}

 Although our analytic results are formally accurate only for $d_p\lesssim$ several km, there is hardly a reason to believe that as they grow larger, planetesimals will become more vulnerable to destruction.  Indeed, larger bodies are more resistant to higher velocity collisions. Also, we do not expect that they will have collision velocities dramatically higher than their lower-$d_p$ predecessors. In fact, it seems much more likely that large bodies will have {\it smaller} collision velocities.  This is because close to the binary, $|A| \approx |A_b| \gg |\dot \varpi_b|$, so $d_c$ (for which $|A|\tau_d\sim 1$, if we ignore $e_\phi$ in Equation \eqref{eq:atd}) is much smaller than the size of bodies for which $\tau_d \dot \varpi_b \sim 1$. Thus, larger bodies should be more resistant to destruction, see Figure \ref{erosionFig}.

%%%%%%%%%%%%%%%%%%%%%%%%%%%%%%%%%%%%%%%%%%%%%
%%%%%%%%%%%%%%%%%%%%%%%%%%%%%%%%%%%%%%%%%%%%%

\section{discussion}  
\label{sect: discussion}

%%%%%%%%%%%%%%%%%%%%%%%%%%%%%%%%%%%%%%%%%%%%%

Here we will revisit some of the assumptions that went into the model described in the previous sections, and explore what happens when they are relaxed.  

%%%%%%%%%%%%%%%%%%%%%%%%%%%%%%%%%%%%%%%%%%%%%

\subsection{Erosion}
\label{sect:erosion}

%%%%%%%%%%%%%%%%%%%%%%%%%%%%%%%%%%%%%%%%%%%%%

In the preceding sections, we have assumed that planetesimal growth is inevitable in the absence of catastrophic disruption events.  This is not necessarily the case, as frequent collisions with smaller objects can still lead to mass loss, even though none of them are severe enough to destroy the planetesimal.  This is the erosion regime defined in Section \ref{sect:CollisionalOutcomes}.  In RS15b (see their \S 4), we determined that in a collision between planetesimals of mass $m_1$ and $m_2$, the critical velocity for erosion is independent of $m_2$ in the limit that $m_2 \ll m_1$.   Erosion can therefore be deleterious in regions where planetesimals are safe from catastrophic disruption because a low value of $d_c$ does not have as large of a protective influence.

If characteristic eccentricity $e_c$ is high enough for collisions to be erosive in the limit of vanishing collision partner mass, then erosion might be expected to inhibit planetesimal growth. Looking at Figure \ref{erosionFig}, we see that erosion is apparently ubiquitous, even in a regime where strong planetesimals do not suffer catastrophic disruption. Therefore, if a substantial fraction of the mass that a planetesimal encounters is in small bodies, it might have difficulty growing.  This outcome could be avoided if the majority of the \textit{mass} of solid material in the disk is contained in objects with sizes larger than $d_{\rm min}$, or if collision rates between large and small bodies are suppressed. 

In practice, we do not expect erosion to be the major obstacle to planetesimal growth because small bodies are likely to be rapidly flushed out of the system due to gas drag in the slightly sub-Keplerian disk.

Indeed, using the results of \citet{Adachi76}, and ignoring the relative particle-gas eccentricity, we estimate an in-spiral timescale of
\begin{align}
\label{eq:tau_d}
& \tau_m  = \frac{a_p}{\dot a_p} = \frac{256 \sqrt{2 \pi}}{507 C_d E(\sqrt{3}/2)} n_p^{-1} \frac{\rho_p d_p}{\Sigma_d}\left(\frac{a_p}{h}\right)^3\\
& \approx 6 \times 10^4{\rm yr} \times \nonumber \\
&\frac{\rho_p}{{\rm 3~ g\; cm^{-3}}} \frac{d_p}{1.3 {\rm m}}  \frac{{\rm 3,\!000 \; g \; cm^{-2}}}{\Sigma_0}\left(\frac{M_b}{0.89M_\odot}\right)^{1.5} a_{p,5}^{9/4}. \nonumber
\end{align}
In this estimate we have taken for $d_p$ the critical size of $d_c\approx 1.3$ m given by Equation \eqref{eq:dc} at 5 AU.  Particles with sizes above $d_c$ are less likely to erode larger bodies due to their aligned orbits.  This is a lower bound, as non-zero gas-particle eccentricity will only accelerate in-spiral.  

Furthermore, as discussed in \citet{Weidenschilling77}, for these small bodies moving slowly relative to the gas, the drag is actually not in a quadratic regime, and is stronger than predicted by our assumed quadratic drag law, making the in-spiral more rapid.   The particles which in-spiral most rapidly are those for which $n_p \tau_d \approx 1$.  Using our quadratic drag law, $n_p \tau_d = 1$ at 5 AU for particles with sizes of about a centimeter.  The fact that we are doing this estimate at 5 AU, instead of the conventional 1 AU, leads to meter sized particles lasting substantially longer than the 100 yr timescale generally discussed in the literature \citep{Weidenschilling77}.  

The significance of the estimate (\ref{eq:tau_d}) is that the disk cannot store most of its mass in bodies smaller than $d_c$ because $\tau_m$ is considerably shorter than the several Myr disk lifetime. As a result, the disk would rapidly lose all of its solid material by radial drift towards the central binary. By assuming most of the mass in the disk to be in planetesimals larger than several meters initially, we thus make erosion by smaller bodies to be a subdominant effect. 

In addition, very small bodies with stopping times shorter than the dynamical time $n_p^{-1}$ orbit with sub-Keplerian velocities due to their strong coupling to gas, which experiences radial pressure support. These objects would impact larger bodies at speeds of several tens of m s$^{-1}$, even in the absence of any secular excitation. The same would happen in protoplanetary disks around single stars, which according to recent statistics of exoplanets, have no diffuculty of forming planets. Thus, erosion by very small objects can be overcome in circumbinary systems in the same way as this happens around single stars.

%%%%%%%%%%%%%%%%%%%%%%%%%%%%%%%%%%%%%%%%%%%%%

\subsection{Growth via ``Lucky" Bodies}

%%%%%%%%%%%%%%%%%%%%%%%%%%%%%%%%%%%%%%%%%%%%%

It is likely that planetesimal growth can occur even in presence of {\it some} catastrophic disruption, since the detrimental effect of these can be offset by other favorable collisions which enable growth.  This balance of the mass loss and gain has been explored for the growth of centimeter-sized bodies by \citet{Windmark12} and \citet{Garaud13}. These authors consider a {\it distribution} of encounter velocities and collision outcomes, rather than assuming all encounters to be at the mean velocity. By using such statistical approach they find over an order of magnitude change in the size of the largest particles that can grow in their coagulation simulations. Thus, statistical nature of coagulation may be an important factor of planetesimal growth, which we did not account for here.

In the km-sized planetesimal regime, bodies generally become both more resistant to collision, {\it and} experience lower velocity collisions with similar sized objects as they grow larger because their size moves further from $d_c$, see Equation \eqref{eq:e12small}.  It therefore seems likely that including the full distribution of collision velocities will make an even larger impact in this scenario than it did for the growth of centimeter sized grains, since a km-sized body that grows via a few lucky collisions becomes harder to destroy. We leave the detailed exploration of the statistical growth of planetesimals in circumbinary systems to future work.

%%%%%%%%%%%%%%%%%%%%%%%%%%%%%%%%%%%%%%%%%%%%%

\subsection{Limits of the Apsidally Aligned Regime}
\label{sect: limits of apsidally aligned}

%%%%%%%%%%%%%%%%%%%%%%%%%%%%%%%%%%%%%%%%%%%%%

In the previous sections we have assumed that free eccentricity of planetesimals has been completely damped by gas drag.  As the free eccentricity goes away on the timescale $\tau_d$, we need to demonstrate that $\tau_d$ is shorter that the characteristic planet formation timescale. 

At a given semi-major axis in the disk, we can use Equation (\ref{taud}) to solve for the critical damping size $d_{\rm damp}$ at which the free eccentricity damping time $\tau_d$ is equal to some characteristic time $\tau$ (not to be confused with the critical size $d_c$ for which the damping time is roughly the precession time $A^{-1}$!). If we set $\tau$ equal to the expected lifetime of the circumbinary disks (Myrs), then all objects with $d_p > d_{\rm damp}$ will not have a chance to settle to their quasi-equilibrium forced eccentricities during the disk evolution. This would violate our basic assumption of damped $e_{\rm free}$ stated in \S \ref{sect:dynamicsgen} and would make planetesimal growth more complicated.

\begin{figure}
\centering
\includegraphics[scale = .5]{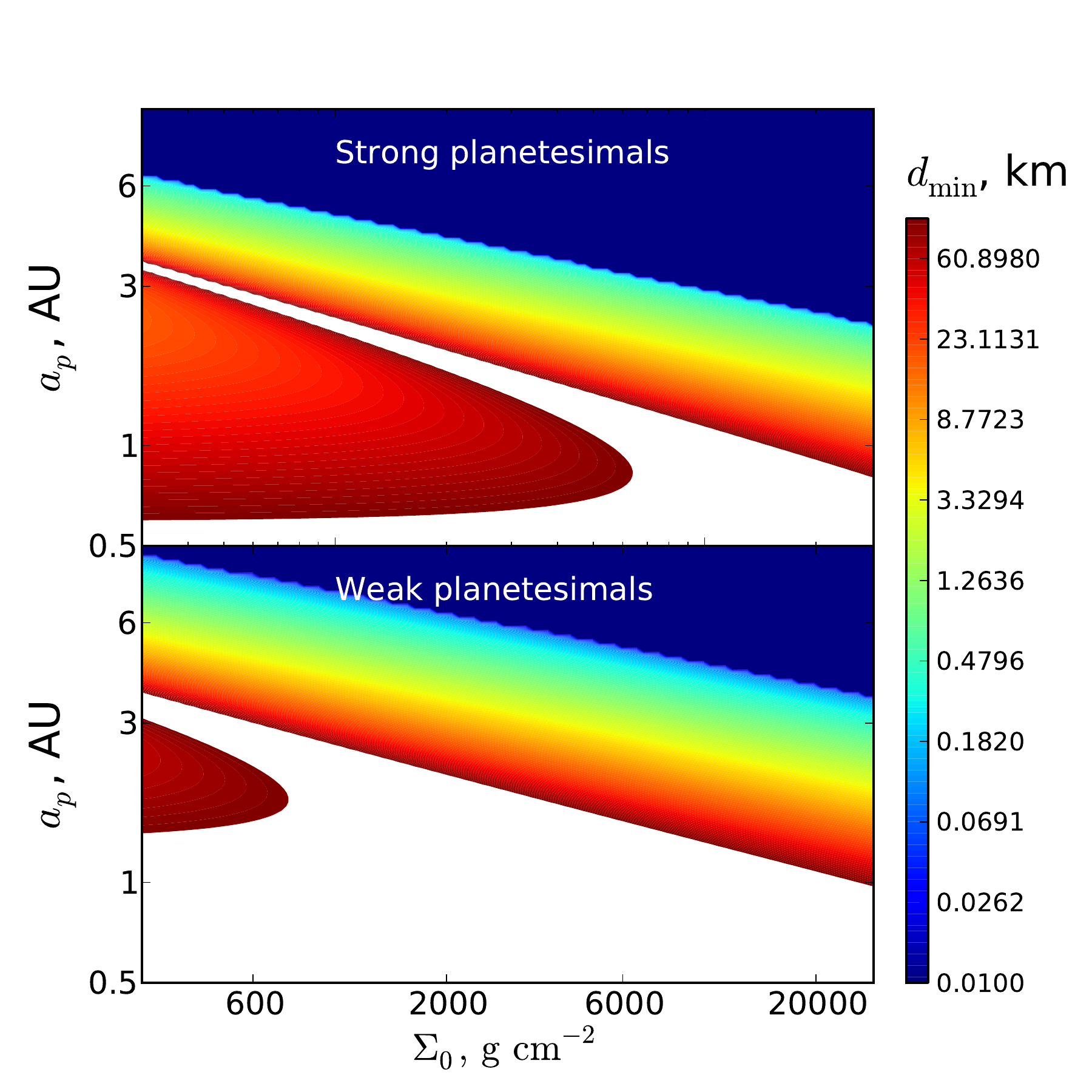}
\caption{$d_{\rm min}$ for the same dynamical environment as Figure \ref{axisymmetricSizePlot}, but assuming planetesimal orbits to be unaligned so that $e_{12} = e_c$.  Comparison of these two figures shows that the apsidal alignment is a very important effect, greatly facilitating planetesimal growth.}
\label{unaligned}
\vspace{-.05cm}
\end{figure}

To demonstrate this last point, we repeat the calculation illustrated in Figure \ref{axisymmetricSizePlot}, however, now we do not assume planetesimal orbits to be aligned according to their forced eccentricity values. Instead, we set $e_{12}$ to be given by $e_c$ instead of by Equation \eqref{eq:e12}.  In the approximation where planetesimals decouple instantly from the gas with eccentricity equal to $e_d$, their eccentricity vector ${\bf e}_p$ circulates around the forced eccentricity with a magnitude $|{\bf e}_d - {\bf e}_{\rm forced}|$.  As shown in RS15a, $|{\bf e}_d - {\bf e}_{\rm forced}|$ is just $e_c$.  Therefore a typical relative eccentricity $e_{12}$ between planetesimals in this population will be of order $e_c$.  This is the approximation explored in \citet{R13}. 

Results of such calculation are shown in Figure \ref{unaligned}. Comparing Figures \ref{axisymmetricSizePlot} and \ref{unaligned} shows that the alignment assumption makes a substantial difference, with the zone of catastrophic disruption significantly extended in Figure \ref{unaligned} compared with Figure \ref{axisymmetricSizePlot} for a given planetesimal size. Thus, lack of free eccentricity damping resulting in apsidal misalignment of planetesimals should have a deleterious effect on planetesimal growth \citep{Kenyon2015}. 

\begin{figure}
\centering
\includegraphics[scale = .3]{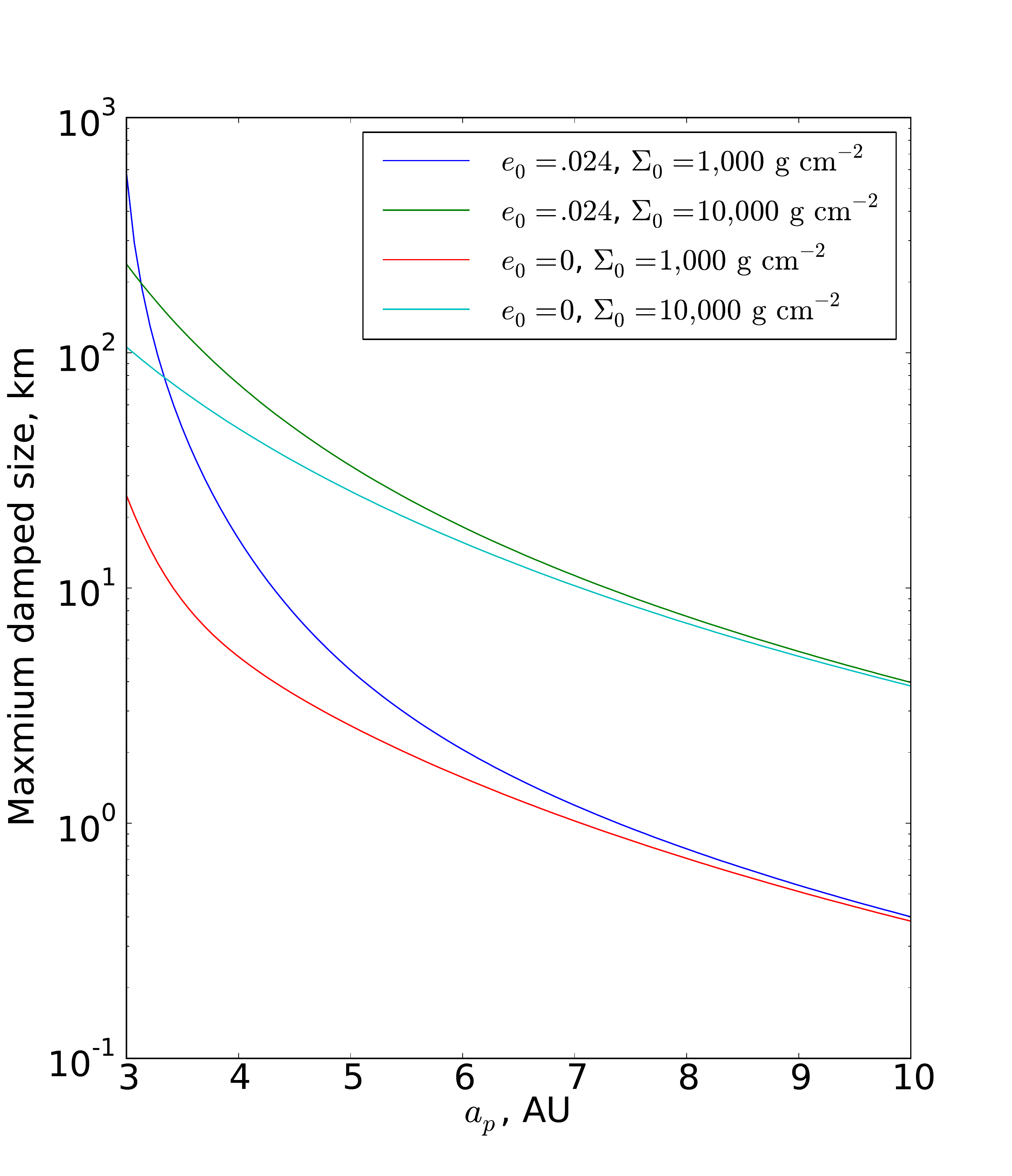}
\caption{Maximum size of planetesimal in the disk dominated excitation regime (where $e_c$ is given by Equation \eqref{ddec}) such that $\tau_d$ is less than 3 Myr.  We have assumed fiducial system parameters except where noted.}
\label{timescales}
\vspace{-.05cm}
\end{figure}

We expect protoplanetary disks to last for a few Myr \citep{Haisch01}. This motivates us to set a characteristic timescale in the determination of the critical size $d_{\rm damp}$ to be $\sim$ Myr. Figure \ref{timescales} shows $d_{\rm damp}$ for $\tau=3$ Myr as a function of $a_p$ for several sets of disk parameters.  This size is calculated numerically from Equations \eqref{eq:sigmae}, \eqref{eq:Ad}, \eqref{AB}, \eqref{eqhoverr}, \eqref{eq:dc}, \eqref{eq:atd}, \eqref{eq:ec}, \eqref{eq:Advarpid},  and \eqref{axisymec}.   

We see that generally within 10 AU, the damping time is shorter than the disk lifetime for kilometer sized planetesimals, depending, of course, on the density of the disk.  Although outside of 5 AU, we may be concerned about the alignment of 10 km planetesimals, at these radii, such large planetesimals are generally well above the mass threshold for collisional growth (see \S \ref{sect:axisymmetricdiskdynamics}).  This suggests that we are justified in considering the planetesimal orbits to be aligned.  It is also worth noting that a higher value of $e_c$ leads to faster apsidal alignment, so the environments with high $e_c$, which present the highest danger to planetesimal coagulation are the same environments which are helped most by the apsidal alignment.  
\par
Collisions between planetesimals may also cause misalignment.  However in the regions of the disk with long alignment timescales, planetesimals greater than $\sim 10$ m in size possess nearly apsidally aligned equilibrium orbits, so collisions between them are unlikely to lead to much misalignment.

%%%%%%%%%%%%%%%%%%%%%%%%%%%%%%%%%%%%%%%%%%%%%

\subsection{Disk Density Structure}

%%%%%%%%%%%%%%%%%%%%%%%%%%%%%%%%%%%%%%%%%%%%%

One may wonder how our results would change if the slopes of the $\Sigma_d(a_p)$ and $e_d(a_p)$ dependencies were varied. Observational evidence for the values of $p$ and $q$ is rather scant and is based predominantly in the sub-mm dust continuum observations on scales (tens of AU) much larger than the ones considered in this work\footnote{{\it ALMA} may be close to marginally resolving these scales in the future.} \citep{Andrews09}. However, the total masses of the circumbinary disks are found to be in the range of several percent of M$_\odot$ \citep{Harris12}, in agreement with our estimate (\ref{eq:M_d}). 

We have picked our fiducial value of $\Sigma_0$ to match an overall disk mass of a few percent of $M_\odot$, believing this to be better constrained than the surface density $\Sigma_0$ at 1 AU.  If we change $p$ while keeping the overall mass of the disk constant, we see that $\Sigma_0$ would vary with $p$ as 
\begin{equation}
\Sigma_0 \propto (2-p) \left(\frac{a_{\rm out}}{a_0}\right)^{p}.
\end{equation}
Equations \eqref{eq:Ad} and \eqref{eq:Bd} show that $\Sigma_d$ and $\psi_1$ are the only disk properties that determine $A_d$.  The coefficients $\psi_1$ and $\psi_2$ in Equations \eqref{eq:Ad} and \eqref{eq:Bd}  change by less than a factor of 2 between $p = 1.5$ and $p = 0.5$, so we see that at a given radius, changing $p$ at constant disk mass has about the same effect as changing $\Sigma_0$, as was done in Section \ref{sect:disk mass outcomes}.  In other words, because $\psi_1$ and $\psi_2$ vary so weakly with $p$ and $q$, what matters most for the disk disturbing function is the {\it local} value of the surface density, not the distribution in the whole disk. And it was shown in SR15 that the edges of the disk are unimportant in determining the disturbing function for the values of $p$ and $q$ that we consider in this paper.  

%%%%%%%%%%%%%%%%%%%%%%%%%%%%%%%%%%%%%%%%%%%%%

\subsection{Non-Secular Terms in the Disturbing Function}

%%%%%%%%%%%%%%%%%%%%%%%%%%%%%%%%%%%%%%%%%%%%%

In this study, we ignore the possibility of mean-motion resonances between the binary and the planetesimals. Previously, \citet{Meschiari2014} found in simulations of the Kepler 16 system that planetesimals got trapped at the 5:1 resonance.  Simulating the S-type $\gamma$ Cephei system, \citet{Leiva13} find planetesimals getting trapped in first order resonances as high as 16:1. While it is not clear if these results would hold far from the binary where gravity of a massive eccentric disk dominates, the mean motion resonances may still play an important role in the dynamics inside of 1 AU. 

It was also suggested in \citet{Paardekooper2012} that short-period terms in the disturbing function, varying with the binary period or the planetesimal orbital period will add additional eccentricity.  These seem unlikely to mis-align aligned orbits: short period terms should affect planetesimals of all sizes equally, not changing their relative velocities, as they act on a much shorter time-scale ($\sim n_p^{-1}$) than the gas drag damping time $\tau_d$ for the sizes we are concerned with.  It was furthermore shown both analytically and in simulation by \citet{Kenyon2015} that in the Kepler 16 system these terms only induce absolute eccentricity variations of about a percent at 0.7 AU (relative variation should be much smaller).

%%%%%%%%%%%%%%%%%%%%%%%%%%%%%%%%%%%%%%%%%%%%%
%%%%%%%%%%%%%%%%%%%%%%%%%%%%%%%%%%%%%%%%%%%%%

\section{Comparison with Previous Work on Planet Formation in Binaries}
\label{sect:comp}

%%%%%%%%%%%%%%%%%%%%%%%%%%%%%%%%%%%%%%%%%%%%%

Here we compare our findings with existing results on planet formation in both circumbinary and circumstellar systems and put them in broader context.  

%%%%%%%%%%%%%%%%%%%%%%%%%%%%%%%%%%%%%%%%%%%%%

\subsection{Previous Work on Circumbinary Planet Formation}
\label{prevWork}

%%%%%%%%%%%%%%%%%%%%%%%%%%%%%%%%%%%%%%%%%%%%%

Despite the different setup and the range of physical effects taken into account in this work, many of our conclusions are similar to those reached in a number of previous studies of planet formation in circumbinary systems.

\citet{Paardekooper2012} and \citet{Meschiari2012} numerically explored the interactions between swarms of planetesimals embedded in an axisymmetric gas disk around a central binary.  They included the gravitational perturbations from the binary, and gas drag from the disk, but not the gravity of the disk, which as we know now (Rafikov 2013, SR15, RS15a) provides the most important gravitational effect outside of a few AU.  They both found it difficult to explain in-situ accretion at $\lesssim 1$ AU separation without invoking initial planetesimals greater than 10 km in size.   

\citet{Meschiari2012} concluded that formation outside of 4 AU was possible starting from km-sized planetesimals, which is only slightly less optimistic than our conclusions for the axisymmetric disk of a similar mass.  Our inclusion of disk gravity in this paper pushes the boundary of the coagulation region only a little bit inward because we find that the secular resonance, emerging when we account for disk gravity, makes conditions in the disk around 2 AU unfavorable for planetesimal growth.

\citet{Meschiari2014} did include disk gravity in some of his simulations and found prominent secular resonance predicted in \citet{R13}. However, he was primarily interested in in-situ growth of circumbinary planets within 1 AU and did not explore planetesimal growth at several AU in as much detail as we do in \S \ref{sect:pgdder}. At $a_p\lesssim 1$ AU \citet{Meschiari2014} finds that km-sized planetesimals can grow by accretion of small collisional debris. By carefully examining planetesimal dynamics in \S \ref{sect:fastStop} we find that growth starting with km-sized objects is possible even with more standard collisional scenario not involving accretion of small particles, as long as the disk density is not very high and the characteristic planetesimal size $d_c$ is small.

Simulations of \citet{Marzari13} also include the gravitational effect of a non-axisymmetric disk.  They calculated the disk structure in 2D geometry, and found complex behavior for disk surface density and eccentricity with radius: their disk maintains substantial $e_d$ (several percent) out to nearly 10 AU. This eccentricity profile looks very different (much higher) than found in simulations of \citet{Meschiari2014}, \citet{Pelupessy13}, and may reflect transient behavior. As a result, \citet{Marzari13} found it difficult to grow even 25 km planetesimals out to 10 AU because of eccentricity excitation due to disk gravity. This is at odds with our findings simply because we think that such eccentricity profiles are unlikely and assume $e_d$ to decay with radius. Another reason for the discrepancy is the duration of their runs ($\sim 10^4$ yr), which is likely shorter that $\tau_d$ for the 5-25 km planetesimals they are considering. This leaves significant undamped free eccentricity and results in large collisional velocities even among planetesimals of the same size.

\citet{Kenyon2015} performed an analytic study of planetesimal dynamics, and showed in particular that short-period terms in the disturbing function do not result in increased collision velocities of planetesimals, in agreement with the argument of \citet{R13}. However, their prime focus was on pointing out the existence of a set of non-crossing, aligned orbits of planetesimals, which they found explicitly neglecting gas drag. They argued that such orbits would allow planetesimals to collide with very low relative speeds (as low as in axisymmetric disks around single stars) and grow efficiently. 

Based on our results, we interpret these trajectories as orbits with fully damped free eccentricity and only forced eccentricity remaining, which by itself requires some effective damping process (most likely gas drag) and works only for objects with $d_p\lesssim 10$ km at several AU, see \S \ref{sect: limits of apsidally aligned}. For such objects gas drag would in fact make forced ${\bf e}_p$ {\it size-dependent} (\S \ref{sect:exactNoDiskPrecess}, \ref{sect:diskdominatedregimedynamics}), leading to substantial collisional velocities for planetesimals of different sizes even when all free eccentricity is damped, as clearly demonstrated in our \S \ref{sect:CollisionalOutcomes} (e.g. see Figure \ref{erosionFig}). This dynamical size segregation was previously broadly discussed in the context of planetesimal growth in S-type systems (e.g. \citet{Thebault2008}, RS15a).  Additionally, when one considers disk gravity, secular resonances arise, leading to orbit crossing even assuming apsidally aligned orbits, due to the rapid radial variation of $e_p$.

%%%%%%%%%%%%%%%%%%%%%%%%%%%%%%%%%%%%%%%%%%%%%

\subsection{Large Initial Planetesimals}
\label{LIPs}
%%%%%%%%%%%%%%%%%%%%%%%%%%%%%%%%%%%%%%%%%%%%%

It is possible that some mechanism other than collisional agglomeration produces a population of large initial planetesimals resistant to collisional destruction by virtue of their size.  It has been proposed \citep{Jeremy2000, YG05} that coupling between solid material and the disk could produce an instability that would lead to overdense rings of material which would collapse to form planetesimals.  This so-called streaming instability has been explored numerically in \citet{Johansen12}, and was found to lead to rapid formation of planetesimals several hundreds of km in size.   

The size distribution of minor bodies in our Solar System provides clues as to the size of the initial planetesimals, however studies of this have come to conflicting results.  \citet{Morbidelli09} find evidence for 100 -1000 km initial planetesimals in the current size distribution of asteroids.  However, \citet{Weidenschilling11}, using a different accretion model, is able to reproduce the current size distribution starting from $<0.1$ km planetesimals.  Additionally,  \citet{Schlichting13} studied the size distribution of objects in the Kuiper belt, and came to the conclusion that initial planetesimals on the order of a km in size provide the best fit.  

Our present work shows that at least in circumbinary systems, rapid formation of large ($d_p\sim 10^2$ km) planetesimals is not necessary and planetesimal growth is possible starting from km-sized (or even smaller, at large separations) bodies. Thus, at present, circumbinary exoplanets do not provide a strong argument in favor of streaming instability being the dominant planetesimal formation mechanism.  It is also unclear whether the streaming instability would function in the perturbed circumbinary environment.

%%%%%%%%%%%%%%%%%%%%%%%%%%%%%%%%%%%%%%%%%%%%%

\subsection{Differences Between P-type and S-type Binaries}
\label{sect:differences}

%%%%%%%%%%%%%%%%%%%%%%%%%%%%%%%%%%%%%%%%%%%%%

This paper has a lot in common with RS15a,b.  In this section we highlight some differences of the  physics that goes into understanding dynamics in P-type systems.

In S-type systems, there is evidence \citep{Muller12} that massive disks end up apsidally aligned with the binary star. This alignment gives rise to a dynamically cold region in the protoplanetary disks of the S-type systems discovered in RS15a, which cannot exist in the circumbinary configuration. Circumbinary disks can in fact contain a dynamically cold region, but only if they are precessing in the prograde sense (see panels C and D of Figure \ref{DiskDominatedErRun}). 

The precession of the central binary cannot be ignored in P-type systems. This helps to lower the planetesimal eccentricity excitation due to the binary. This effect is absent in S-type systems.

Circumbinary disks can be much more extended than circumprimary ones, simply because they are not tidally truncated by the companion on the outside. For that reason planet formation can (and is likely to) occur at large radii (with subsequent inward migration of grown planets).  We see from Equation \eqref{eq:dc} that everything else being equal, the critical size $d_c$ at which collision velocities between similar sized planetesimals are highest scales as $d_c \sim a_p^{-13/4}$.  The low value of $d_c$ in the planet-forming region leads to the conclusion that all bodies km-sized or larger are very nearly apsidally aligned as $d_p \gg d_c$ for all of them, see Equation \eqref{eq:e12small}.  Additionally, the Keplerian velocity is smaller at higher separation, leading to further reduction of collision velocities. This permits planetesimal coagulation starting from relatively small initial sizes, 10 - 100 m, even in regions of higher $e_c$ than is permitted in the S-type systems where massive planets reside at 1-2.5 AU.

Another consequence of the low density environment at large separation in circumbinary systems is that for a substantial fraction of the likely planet-forming part of the disk, there is some concern that alignment with the gas may not occur within the disk lifetime, see Section \ref{sect: limits of apsidally aligned}. This was not an issue for the S-type systems because their compact protoplanetary disks were likely much denser around 1-2 AU, effectively damping free eccentricity and aligning planetesimal orbits.

%%%%%%%%%%%%%%%%%%%%%%%%%%%%%%%%%%%%%%%%%%%%%
%%%%%%%%%%%%%%%%%%%%%%%%%%%%%%%%%%%%%%%%%%%%%

\section{Summary}
\label{sect:summary}

%%%%%%%%%%%%%%%%%%%%%%%%%%%%%%%%%%%%%%%%%%%%%

We studied planetesimal dynamics in circumbinary disks with the goal of understanding the conditions leading to planetesimal growth, which is a natural step towards formation of circumbinary planets such as Kepler-16.  Our study simultaneously considers (1) gas drag (including non-trivial radial pressure support in the gaseous disk), (2) gravity from the eccentric precessing binary, and (3) gravity from the eccentric precessing disk. We found analytical solutions for planetesimal eccentricity behavior in many important limits. 

We estimate the precession rate of the central binary and find binary precession to play a non-trivial role in the determination of planetesimal dynamics. We believe erosion by small bodies to not be a major hindrance to growth, as such objects should spiral in towards the central binary on timecales of tens of thousands of years.  
Based on our analytical results, we make the following conclusions:

\begin{itemize}

\item We find disk gravity to play the dominant role in planetesimal dynamics outside of several AU. Secular resonances (up to three for some choices of the disk and binary precession rates) significantly complicate planetesimal dynamics at $a_p\sim$ several AU.

\item Apsidal precession of the central binary reduces the direct effect of binary gravity on planetesimal orbits. For many circumbinary systems discovered by {\it Kepler} precession is dominated by the gravity of the protoplanetary disk.

\item If planet formation is precluded only by catastrophic disruption, then we find that forming Kepler-16 and similar planets in situ requires that initial planetesimal sizes were large and disk masses were small. For example, even in the most favorable case of an axisymmetric disk with a mass of only a few Jupiter masses, we still require 1 km initial planetesimals to avoid catastrophic disruption at 0.7 AU.  If the disk were eccentric or more massive, then even larger initial planetesimals would be required.

\item Formation outside of $\sim 3$ AU is much easier and more likely. Here a dense ($\Sigma_d(1{\rm AU}) > 10^4$ g cm$^{-2}$) disk (in agreement with sub-mm observations) is helpful as its gravity moves the secular resonance inwards of 2 AU, lowering planetesimal-planetesimal collision velocity.  The dense disk also provides enough eccentricity damping to align $\sim 10$ km-sized planetesimals out to 10 AU during the disk lifetime.

\item Disk precession can either facilitate or hinder plantesimal coagulation, depending on the direction and magnitude of precession. Slow retrograde precession results in emergence of a destructive secular resonance at several AU.  On the other hand, prograde precession leads to a dynamically favorable location in the disk where the forced eccentricity is the same as the local gas eccentricity, relative planetesimal eccentricities are small, and there is no dynamical barrier to coagulation.

\end{itemize}

This work thus provides a dynamically-motivated picture of planetesimal growth towards building planetary cores in circumbinary protoplanetary disks.

%%%%%%%%%%%%%%%%%%%%%%%%%%%%%%%%%%%%%%%%%%%%%

\acknowledgements

This work is supported by NSF grant AST-1409524.

%%%%%%%%%%%%%%%%%%%%%%%%%%%%%%%%%%%%%%%%%%%%%

\bibliographystyle{apj}
\bibliography{apj-jour,ms.bib}
\appendix
\numberwithin{equation}{section}

%%%%%%%%%%%%%%%%%%%%%%%%%%%%%%%%%%%%%%%%%%%%%
%%%%%%%%%%%%%%%%%%%%%%%%%%%%%%%%%%%%%%%%%%%%%

\section{Planetesimal dynamics with linear drag}
\label{sect:lindrag}

%%%%%%%%%%%%%%%%%%%%%%%%%%%%%%%%%%%%%%%%%%%%%

In the case of linear gas drag (i.e. $\tau_d$ independent of $e_r$) we can obtain the complete analytical solution for planetesimal dynamics including the precession of both the binary and the disk. With fixed $\tau_d$ and $\varpi_b(t)=\dot \varpi_b t$, $\varpi_d(t)=\dot \varpi_d t$ our general evolution equations (\ref{DDEquations1})-(\ref{DDEquations2}) admit the following analytical solution:
\begin{eqnarray}
&  \left\{
\begin{array}{l}
k_p\\
h_p
\end{array}
\right\}
 =e_{\rm free}e^{-t/\tau_d}
\left\{
\begin{array}{l}
\cos\left(At+\varpi_0\right)\\
\sin\left(At+\varpi_0\right)
\end{array}
\right\}
+ 
\left\{
\begin{array}{l}
k_{f,d}\\
h_{f,d}
\end{array}
\right\}
+ 
\left\{
\begin{array}{l}
k_{f,b}\\
h_{f,b}
\end{array}
\right\}
.
\end{eqnarray}
Here $e_{\rm free}$ and $\varpi_0$ are the free eccentricity and periastron angle, $k_{f, d}$ and $h_{f, d}$ are the components of the forced eccentricity associated with the disk, and $k_{f, b}$, and $h_{f, b}$ are the components associated with the binary. These are given by
\begin{eqnarray}
&  \left\{
\begin{array}{l}
k_{f,d}\\
h_{f,d}
\end{array}
\right\}
 = \left[\frac{e_g^2 + \tau_d^2 B_d^2}{1 + \tau_d^2(A - \dot \varpi_d)^2} \right]^{1/2}
 \left\{
\begin{array}{l}
\cos{(\varpi_d(t) + \phi_d)}\\
\sin{(\varpi_d(t) + \phi_d)} 
\end{array}
\right\}
, \quad
\cos{\phi_d} = \frac{e_g - \tau_d^2B_d(A - \dot \varpi_d)}{(e_g^2 + \tau_d^2 B_d^2)^{1/2} \left[1 + \tau_d^2 (A - \dot \varpi_d)^2\right]^{1/2}},
\end{eqnarray}
and 
\begin{eqnarray}
& \left\{
\begin{array}{l}
k_{f,b}\\
h_{f,b}
\end{array}
\right\}
 = \left[\frac{B_b^2 \tau_d^2}{1 + \tau_d^2(A - \dot \varpi_b)^2}\right]^{1/2}
 \left\{
\begin{array}{l}
\cos{(\varpi_b(t) + \phi_b)}\\
\sin{(\varpi_b(t) + \phi_b)} 
\end{array}
\right\}
, \quad
\cos{\phi_b} = \frac{-(A-\dot \varpi_b) \tau_d}{ \left[1 + \tau_d^2 (A - \dot \varpi_b)^2\right]^{1/2}}.
\end{eqnarray}
The relative particle-gas eccentricity is given by 
\begin{eqnarray}
&\left\{
\begin{array}{l}
k_{r}\\
h_{r}
\end{array}
\right\}
 = 
 \left\{
\begin{array}{l}
k_{f,b}\\
h_{f,b}
\end{array}
\right\}
- \tau_d \frac{B_d + e_g(A - \dot \varpi_d)}{\left[1 + \tau_d^2(A - \dot \varpi_d)^2 \right]^{1/2}}
\left\{
\begin{array}{l}
\cos{(\varpi_d(t) - \phi_r)}\\
\sin{(\varpi_d(t) - \phi_r)}
\end{array}
\right\}
, 
\quad
\cos{\phi_r} = \frac{\tau_d(A - \dot \varpi_d)}{\left[1 + \tau_d^2(A-\dot \varpi_d)^2 \right]^{1/2}}.
\end{eqnarray}
It is clear that $e_r$ is in general a function of time.

%%%%%%%%%%%%%%%%%%%%%%%%%%%%%%%%%%%%%%%%%%%%%
%%%%%%%%%%%%%%%%%%%%%%%%%%%%%%%%%%%%%%%%%%%%%

\section{binary precession rate}
\label{sectbinprecess}

%%%%%%%%%%%%%%%%%%%%%%%%%%%%%%%%%%%%%%%%%%%%%

Here we summarize results on the four major causes of binary precession which were discussed in Section \ref{sect:binaryPrecession}.  We write the precession rate as $\dot \varpi_b = \dot \varpi_{GR} + \Sigma_{j = 1}^{2} (\dot \varpi_{T, j}  +  \dot \varpi_{R, j} )+ \dot \varpi_{\rm disk}$, where $\dot \varpi_{GR}$, $\dot \varpi_{T, j}$,  $\dot \varpi_{R, j}$, $\dot \varpi_{\rm disk}$ are the precession rates due to general relativity, tidal and rotational stellar quadrupoles, and disk gravity, respectively.

%%%%%%%%%%%%%%%%%%%%%%%%%%%%%%%%%%%%%%%%%%%%%

\subsection{General Relativistic Precession}

%%%%%%%%%%%%%%%%%%%%%%%%%%%%%%%%%%%%%%%%%%%%%

To first order in eccentricity, precession of Keplerian orbits due to general relativity is \citep{MTW}
\begin{equation}
\dot \varpi_{GR} =  \frac{3(G M_b)^{1.5}}{c^2a_b^{2.5}} = 6.9\times10^{-6} {\rm yr}^{-1}\left(\frac{M_b}{0.89M_\odot}\right)^{1.5} \left(\frac{0.22AU}{a_b}\right)^{2.5}.
\end{equation}

%%%%%%%%%%%%%%%%%%%%%%%%%%%%%%%%%%%%%%%%%%%%%

\subsection{Precession Due to Stellar Quadrupoles Induced by Tides and Rotation}

%%%%%%%%%%%%%%%%%%%%%%%%%%%%%%%%%%%%%%%%%%%%%

Precession rates due to quadrupoles induced by tidal forces and from the rotational bulge are given in \citet {St39} and \citet{Shak85}.  They are given by 
\begin{align}
\label{tpr}
&\dot \varpi_{T, j} = 15 k_{2, j} n_b \frac{M_r}{M_j} \left(\frac{R_j}{a_b}\right)^5 
\nonumber  \\
& = 4.7\times 10^{-6} {\rm yr}^{-1} \frac{k_2}{0.13} \left(\frac{M_b}{0.89M_{\odot}}\right)^{0.5}  \left(\frac{a_b}{0.22AU}\right)^{-6.5} \left(\frac{R_j}{2.03R_\odot}\right)^5,
\end{align}
\begin{align}
\label{rpr}
& \dot \varpi_{R, j} = k_{2, j} \frac{M_b}{M_j} \frac{\omega_j^2}{n_b} \left(\frac{R_j}{a}\right)^5 =
\nonumber  \\
& 1.4 \times 10^{-6} {\rm yr}^{-1}\frac{k_2}{0.13} \left(\frac{M_b}{0.89M_{\odot}}\right)^{0.5}  \left(\frac{a_b}{0.22AU}\right)^{-6.5}  \left(\frac{R_j}{2.03R_\odot}\right)^5,
\end{align}
where $n_b$ is the Keplerian frequency, $r \neq j$, $\omega_j$ is the spin frequency of star $j$, $M_j$ is the mass of star $j$.  $R_j$ is the radius of star $j$, and $k_2$ is the apsidal motion constant, both estimated for the primary in the Kepler 16 system from the pre-main-sequence stellar models of \citet{Cl12} assuming an age of 1 Myr and metallicity of $Z = 0.02$.  For the numerical estimate in Equation \eqref{rpr}, we have assumed the stars to be tidally locked, i.e. $\omega_j =n_b$.  The numerical estimates in Equations \eqref{tpr} and \eqref{rpr} were done assuming the subscript ``$j$'' to refer to the primary star of the Kepler 16 system and the subscript ``$r$'' to refer to the secondary.

\begin{figure}
\centering
\includegraphics[scale = .38]{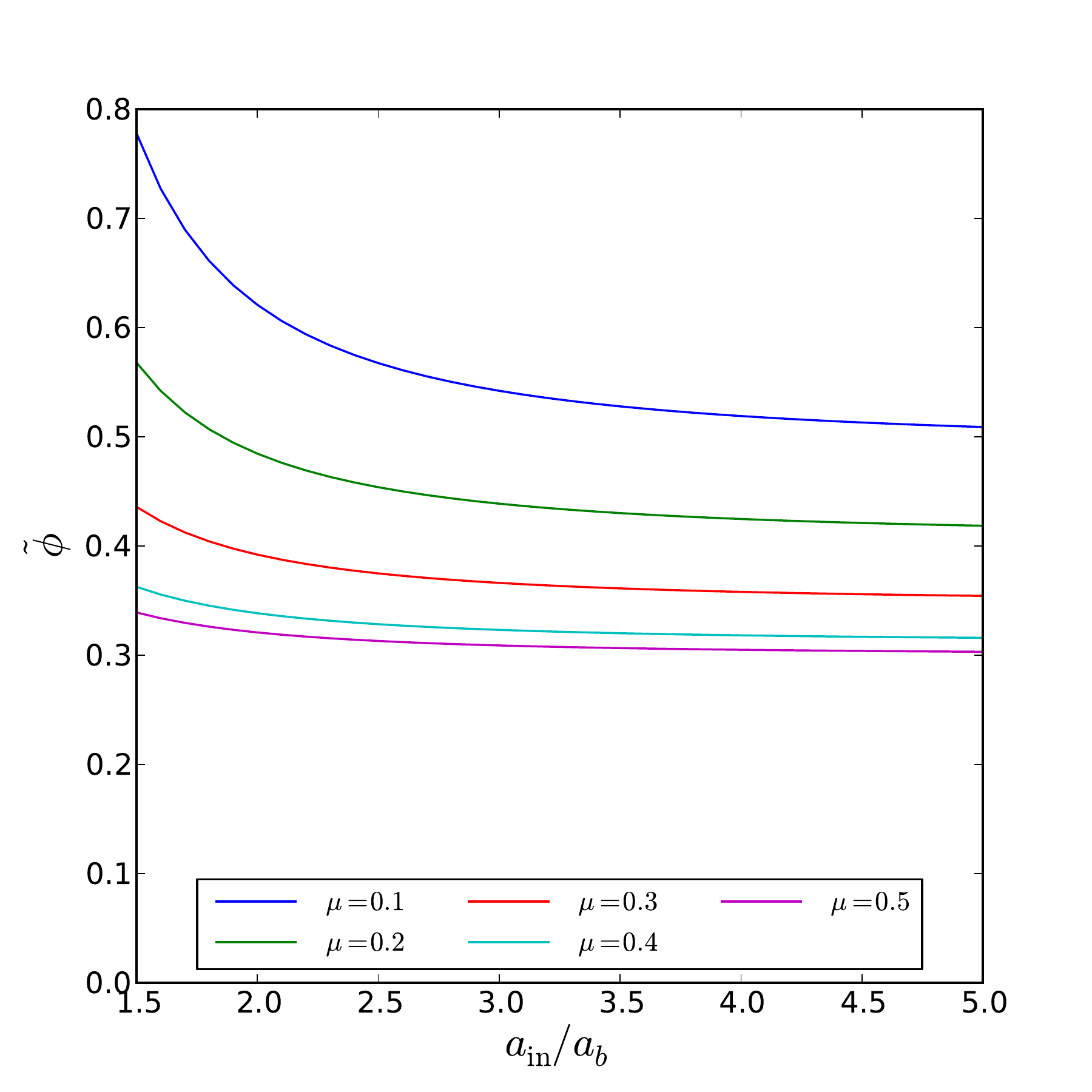}
\caption{Dependence of the dimensionless factor $\tilde \phi(a_b/a_{\rm in}, \mu)$ in the disk-driven binary precession rate $\dot \varpi_{\rm disk}$ (see Equation \eqref{binaryPrecessionFromDisk}) on the relative size of the inner cavity $a_{\rm in}/a_b$ and binary mass ratio $\mu$.}
\label{fig:Phi}
\vspace{-.05cm}
\end{figure}

%%%%%%%%%%%%%%%%%%%%%%%%%%%%%%%%%%%%%%%%%%%%%

\subsection{Precession Due to Disk Gravity}
\label{sect:diskPrecession}

%%%%%%%%%%%%%%%%%%%%%%%%%%%%%%%%%%%%%%%%%%%%%

To calculate the binary precession due to disk gravity, we use Equations (20) and (A3) from \citet{R13}, which assume that power law dependence of the disk surface density is sharply truncated at the inner radius $a_{\rm in}$.  These state 
\begin{align}
\label{binaryPrecessionFromDisk}
&\dot \varpi_{\rm disk} = \pi \tilde \phi(a_b/a_{\rm in}, \mu) n_b\frac{ \Sigma_0 a_0^p a_b^3 }{M_b a_{\rm in}^{1+p}} \nonumber\\
& \approx 2.6 \times 10^{-3} {\rm yr}^{-1} \left(\frac{0.89 M_\odot}{M_b}\right)^{0.5} \left( \frac{a_b}{0.22} \right)^{1/2 - p} \frac{\Sigma_0}{3,\!000 \; \rm{g\; cm^{-2}}} \frac{\tilde \phi}{0.46}.
\end{align}
Here $\tilde \phi(a_b/a_{\rm in})$ is a function of the binary mass ratio $\mu$ and ratio of binary semi-major axis to the radius of the central gap in the disk $a_b/a_{\rm in}$, and we have assumed $a_b/a_{\rm in} = 0.5$. Figure \ref{fig:Phi} shows that $\tilde \phi$ is a slowly varying function of $a_b/a_{\rm in}$, so roughly $\dot \varpi_{\rm disk} \sim (a_b/a_{\rm in})^{1+p}$. For $p = 1.5$, assuming $a_b/a_{\rm in}$ to be 1/3 instead of 1/2 would lower $\dot \varpi_b$ by almost a factor of 3.

%%%%%%%%%%%%%%%%%%%%%%%%%%%%%%%%%%%%%%%%%%%%%
%%%%%%%%%%%%%%%%%%%%%%%%%%%%%%%%%%%%%%%%%%%%%

\section{Collisional Outcomes}
\label{sect: collOut}

%%%%%%%%%%%%%%%%%%%%%%%%%%%%%%%%%%%%%%%%%%%%%

Consider two bodies of mass $m_1$ and $m_2$, colliding at speed $v_{\rm coll}$.  \citet{SL09} give the mass of the largest remnant $M_{\rm lr}$ as
\begin{equation}
\label{eq:SL09}
\frac{M_{\rm lr}}{M_{\rm tot}} = -0.5(Q_R/Q^*_{RD} -1) + 0.5
\end{equation}
Here, $M_{\rm tot} = m_1 + m_2$ is the total mass of the colliding bodies, $Q_R=0.5 m_1 m_2 v_{\rm coll}^2/M_{\rm tot}^2$ is the center of mass specific kinetic energy, and $Q^*_{RD}$ is a quantity dependent on the material properties and $M_{\rm tot}$.  As shown in Figure 2 of RS15b, the critical velocities for catastrophic disruption are on the order of several m s$^{-1}$ for $d_p = 100$ m planetesimals.  The critical speed reflects both the strength of the body, and reaccumulation of fragments due to gravity.  Material strength becomes subdominant to gravity for bodies larger than a few hundred meters.
\par
We take collision velocity to be given by 
\begin{equation}
\label{eq:vcoll}
v_{\rm coll} = \sqrt{(e_{12} v_K)^2 + \frac{2G(m_1 +m_2)}{d_1 + d_2}},
\end{equation}
where the first term is due to the relative velocity at infinity, $e_{12}=|{\bf e}_p(d_1)-{\bf e}_p(d_2)|$, and the second is due to the potential energy of the colliding objects.  It should be noted (RS15a) that the actual velocity at infinity between two colliding bodies can be anywhere between $0.5 e_{12}v_K$ and $e_{12}v_K$.  Thus Equation (\ref{eq:vcoll}) may overestimate the true $v_{\rm coll}$ by up to a factor of 2.  Because the collision velocity must be several times the escape velocity in order to be destructive, the potential energy term is relatively unimportant.  

We can express the relative eccentricity of two planetesimals in terms of $e_c$, $d_c$ and the planetesimal sizes $d_1$ and $d_2$ as (RS15a)
\begin{equation}
\label{eq:e12}
e_{12} = e_c \frac{|(A - \dot \varpi_d) (\tau_1 - \tau_2)|}{\sqrt{(1+(A - \dot \varpi_d)^2 \tau_1^2)(1+(A - \dot \varpi_d)^2 \tau_2^2)}}.
\end{equation}
Here, $|(A - \dot \varpi_d) \tau_i|$ ($i = 1, 2$) is given in terms of $d_c$ and $d_i$ by Equation \eqref{eq:atd}.
\par
In the commonly encountered limit of $d_1$ and $d_2$ both being much greater than $d_c$, this simplifies to
\begin{equation}
\label{eq:e12small}
e_{12} \approx e_c \frac{|d_1 - d_2|d_c}{d_1d_2}.
\end{equation}
We see from this that for encounters between two large bodies, the relative eccentricity is never larger than $e_c d_c/{\rm min}(d_1, d_2)$.

\end{document}